\documentclass[a4paper,cleveref, autoref,thm-restate,numberwithinsect,]{lipics-v2021}
\hypersetup{hypertexnames=false}

\title{Positivity Proofs for Third-Order Holonomic Sequences via Differences Scaled by Rational Functions} 

\titlerunning{Positivity Proofs for 3rd-Order Holonomic Sequences} 

\author{Fugen Hagihara}{
Graduate School of Science, Kyoto University, Japan \url{https://sites.google.com/view/hagihara-fugen/english?authuser=0} }{hagihara@kurims.kyoto-u.ac.jp}{https://orcid.org/0009-0007-0928-7297}{}

\authorrunning{F.~Hagihara} 
\Copyright{Fugen Hagihara} 



\category{} 

\relatedversion{}

\funding{This work was supported by JSPS KAKENHI Grant Number JP25KJ1559.}

\acknowledgements{I would like to thank my supervisor, Akitoshi Kawamura. 
The idea to use the inequality~\eqref{eq:S>=1+P/S+Q/SS} in the proof of \cref{thm:NonPositive} was originally introduced by him during our previous research~\cite{HK25}.
}

\fontsize{11pt}{11pt}\selectfont

\usepackage{amsmath,amssymb,amsthm}
\usepackage{tikz}

\usepackage[all]{xy}
\usepackage{array}
\labelformat{enumi}{#1}
\labelformat{enumii}{\theenumi-#1}

\newcommand{\R}{\mathbb{R}}
\newcommand{\N}{\mathbb{N}}
\newcommand{\Q}{\mathbb{Q}}
\newcommand{\Z}{\mathbb{Z}}

\DeclareMathOperator{\re}{Re}
\DeclareMathOperator{\im}{Im}
\DeclareMathOperator{\cert}{Certificate}

\usepackage{algorithm}%
\usepackage[noend]{algpseudocode}%
\algnewcommand\algorithmicinput{{\bfseries Input}}%
\algnewcommand\algorithmicoutput{{\bfseries Output}}%
\algnewcommand\AlgInput{\item[\algorithmicinput]}%
\algnewcommand\AlgOutput{\item[\algorithmicoutput]}%
\makeatletter
\AtBeginDocument{%
    \captionsetup[algorithm]{name=Procedure}%
}
\makeatother

\crefname{algorithm}{Procedure}{Procedure}

\hideLIPIcs
\begin{document}
\nolinenumbers

\maketitle

\begin{abstract}
We study the positivity of sequences $f$ that satisfy a third-order holonomic recurrence
$f(n+3)= P_2(n) f(n+2) + P_1(n) f(n+1) + P_0(n) f(n)$
whose coefficients $P_0$, $P_1$ and $P_2$ are rational functions in $\R(x)$.
We provide a sufficient condition on $P_0$, $P_1$ and $P_2$ for the recurrence to admit a solution all but finitely many terms of which are positive, and whose positivity is robust under small perturbations of the initial values. 
This sufficient condition is also a necessary condition for a subclass of the recurrences.
By developing the proof of this sufficient condition, for another subclass of the recurrences, we show that it is decidable whether all the terms of a given sequence are positive for any initial values outside an exceptional plane.

It is common to construct a positivity certificate, which guarantees the positivity of a solution, 
to show this kind of decidability. 
In this paper, we construct a new type of positivity certificate via the $S$-scaled difference $f(n+1) - S(n) f(n)$ for a suitable $S \in \R(x)$.
This result partially improves the work by Ibrahim and Salvy.
\end{abstract}

\section{Introduction}
\label{sec:intro}

Let $\N = \{ 0, 1, 2, \dots \}$. Let $r$ be a positive integer. A real-valued sequence $f = \{ f(n) \}_{n \in \N }$ is called an \emph{$r$th-order holonomic sequence} if it satisfies a linear recurrence
\begin{equation}\label{eq:holonomic}
f(n + r) =  P_{r-1}(n) f(n+r-1) + \dots + P_0(n) f(n)  
\end{equation}
for some rational functions $P_0, \dots, P_{r-1} \in \R(x)$.
$f$ is called a \emph{solution} of this recurrence.
When it comes to the decision problems involving holonomic sequences, the sequences are usually assumed to satisfy the recurrence for all $n \in \N$ since shifting the sequences by finitely many terms does not affect the decidability.
We follow this convention.
For the well-definedness of the recurrence, we assume that $\infty \notin P_0(\N), \dots, P_{r-1}(\N)$, i.e., the coefficients have no poles (singular points) in $\N$.

Holonomic sequences arise in various areas of mathematics and computer science. For example, formal power series solutions of linear differential equations with polynomial coefficients are
generating functions of holonomic sequences \cite{Sta80},~\cite[Appendix B.4]{FS13}, and for terms
$F(n, k)$ consisting of binomial coefficients $\binom{n}{k}$
and so on, the sum $f(n) = \sum_{k \in \Z}F(n,k)$ is often holonomic \cite{PWZ96}.
Holonomic sequences also arise in the verification of loops~\cite{HJK17,HJK18}.

In this paper we focus on the positivity of holonomic sequences:

\begin{definition}[Positivity and Ultimate positivity]
    A real-valued sequence $f$ is \emph{positive} (resp. \emph{ultimately positive}) if $f(n) > 0$ for all $n$ (resp. for all sufficiently large $n$). 
\end{definition}

Positivity is sometimes defined as $f(n) \geq 0$ for all $n$.
No matter which style we adopt, we can reconstruct most of the results available now.
The \emph{Positivity Problem} (resp. the \emph{Ultimate Positivity Problem}) asks whether the holonomic sequence defined by given initial values and coefficients of the recurrence is positive (resp. ultimately positive).
It is not known whether these problems are decidable,
while partial results are available.
These problems and partial results are applied to automated inequality proving \cite{GK05,GK06} (see also \cref{sec:execution}) and even to theoretical biology \cite{MM22,BY22,YC22}.

The \emph{Skolem Problem}, asking whether $f(n) = 0$ for some $n$, reduces to the Positivity Problem. (In this reduction, we use a discussion similar to that in \cite[\S~4]{OW12}.)
While the Skolem Problem has been investigated for almost 90 years~\cite{HHHK05},
its decidability is still open. 

It is necessary to restrict to a subfield of $\R$ when we input values into a procedure.
We adopt the field of real algebraic numbers $\overline{\Q} \cap \R$.

\subsection{Previous research}
\label{sec:previous_research}

\subsubsection{Conditions for the existence of an ultimately positive solution}
\label{sec:previous_research:positive_solution}

For general-order holonomic recurrences, \cref{thm:simple_unique_posi_sol} is widely known as a condition for the existence of an ultimately positive solution.
This is obtained from, for example, Ibrahim and Salvy's result~\cite{IS26}.
A recurrence~\eqref{eq:holonomic} is said to be of \emph{Poincar\'{e} type} if each coefficient $P_i(x)$ converges to a real number $p^{(i)}$ as $x \to \infty$. Its \emph{dominant roots} are the roots of maximal modulus of the \emph{characteristic polynomial}
\begin{equation} \label{eq:characteristic_polynomial}
	\chi(x) := x^r - p^{(r-1)} x^{r-1} - \dots - p^{(1)} x - p^{(0)}.
\end{equation}
Note that the restriction to a recurrence of Poincar\'{e} type does not lose generality (see \cref{sec:transformation}).

\begin{theorem} \label{thm:simple_unique_posi_sol}
    Assume that the recurrence~\eqref{eq:holonomic} is of Poincar\'{e} type and has a unique simple dominant root $\lambda \neq 0$ and $0 \notin P_0(\N)$.
    Then the existence of a \emph{robustly} ultimately positive solution $f$ is equivalent to $\lambda>0$,
    where, by robustness, we mean that the ultimate positivity of $f$ is preserved under small perturbations of the initial values.
\end{theorem}

By Ibrahim's result~\cite{Ibr25}, 
for the recurrences with several dominant roots,
there exists a robustly ultimately positive solution if each dominant root is simple and the recurrence satisfies an additional condition.
Sufficient conditions and necessary conditions for third-order recurrences
\begin{equation} \label{eq:general_3rd-order}
f(n+3) = P_2(n) f(n+2) + P_1(n) f(n+1) + P_0(n) f(n)
\end{equation}
to admit an ultimately positive solution are also studied \cite{HP94,PP08,PT02,Par11}.

\subsubsection{Procedures for proving positivity}
\label{sec:positivity_proving_procedure}

The first procedure for proving positivity was proposed by Gerhold and Kauers~\cite{GK05}.
The procedure and when it halts were studied thereafter \cite{KP10,Pil13}.
One of the most recent results is the following. 
Due to the transformation in \cref{sec:transformation}, we can assume that the recurrence is of Poincar\'{e} type.
Note that the statement on the termination below is slightly stronger than the original one. This follows from the construction of the procedure in \cite{IS26}.

\begin{theorem}[Ibrahim, Salvy~\cite{IS26}]\label{thm:simple_unique_procedure}
There exists a procedure for proving positivity of holonomic sequences
which takes as input the coefficients $(P_0, \dots, P_{r-1}) \in (\overline{\Q} \cap \R)(x)^r$
satisfying the condition below
and any initial values $(f(0), \dots, f(r-1)) \in  (\overline{\Q} \cap \R)^r$, and 
has a characterization of termination below.
	\begin{description}
		\item[Coefficient condition]
        The recurrence~\eqref{eq:holonomic}
        is of Poincar\'{e} type and has a unique simple dominant root, and $0 \notin P_0(\N)$.
        \item[A characterization of termination]
		The procedure halts if and only if $(f(0), \dots, f(r-1)) \notin H(P_0, \dots, P_{r-1})$ or the holonomic sequence defined by the input is not positive, where $H(P_0, \dots, P_{r-1}) \subseteq \R^r$ is an $(r-1)$-dimensional subspace depending on the coefficients $(P_0, \dots, P_{r-1})$.
	\end{description}
\end{theorem}

In other words, the Positivity Problem is decidable under the coefficient condition and the characterization of termination, and the characterization of termination is semi-decidable.
While this procedure tries to prove the positivity, it seeks $n$ such that $f(n) \leq 0$ in parallel, and so it halts on any input that does not yield a positive solution.
Note that a similar procedure for proving ultimate positivity can be obtained by slightly modifying the one in~\cite{IS26}, while the characterization of termination will change into $(f(0), \dots, f(r-1)) \notin H(P_0, \dots, P_{r-1})$.

\begin{figure}[htb]
\centering
\includegraphics[width=13cm]{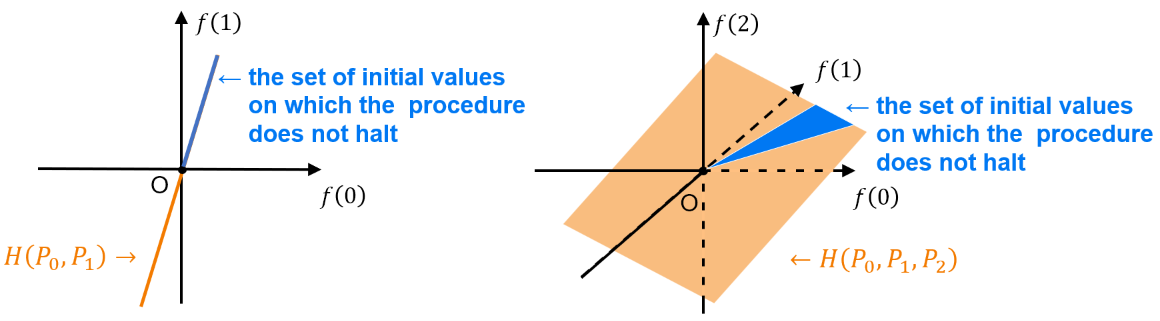}
\caption{Examples of exceptional hyperplanes and the sets of initial values on which the procedure does not halt (left: $r=2$, right: $r=3$).}
\label{fig:}
\end{figure}

The $(r-1)$-dimensional subspace $H(P_0, \dots, P_{r-1})$ is called the \emph{exceptional hyperplane}.
Under the coefficient condition, most initial values are outside the exceptional hyperplane and so the procedure halts on them (\cref{fig:}).
The blue set in the figure represents the set of initial values on which the procedure does not halt, or equivalently, $H(P_0, \dots, P_{r-1}) $ $\cap $ ``the set of initial values which lead to a positive solution.''

If the unique and simple dominant root is negative, by \cref{thm:simple_unique_posi_sol},
there is no robustly ultimately positive solution. 
Then the set of initial values leading to an ultimately positive solution is covered by an $(r-1)$-dimensional subspace.
Hence, in this case, the trivial procedure that only seeks a non-positive term and only outputs ``No'' has a similar halting condition.
Ibrahim~\cite{Ibr25} partially deals with the case with several simple dominant roots.

For second-order holonomic sequences,
Hagihara and Kawamura~\cite{HK25}
showed a similar theorem whose coefficient condition is only $0 \notin P_0 ( \N )$
based on the work~\cite{NOW21} by Neumann, Ouaknine, Worrell and the work~\cite{KKL+21} by Kenison et al.
For third-order holonomic sequences, the state of the art was \cref{thm:simple_unique_procedure} and its extension mentioned above.

Algorithms for positivity and ultimate positivity that are applicable only to relatively narrow classes of recurrences but guaranteed to halt on all initial values are also studied~\cite{NOW21,HK25,Ken24}.

\subsection{Contributions}

The present paper has two lines of contributions: one is on the structure of third-order holonomic recurrences, and the other is on a procedure for proving positivity. 

First, we found a sufficient condition for the third-order recurrences~\eqref{eq:general_3rd-order} to admit a robustly ultimately positive solution, with an explicit boundary $R_P$ (\cref{thm:positive_solution}). 
This condition partially includes the case with a multiple dominant root, so our theorem extends \cref{thm:simple_unique_posi_sol}.
This condition is also a necessary condition in some cases (\cref{thm:NonPositive}), and so $R_P$ is a critical threshold in these cases.
In the sense that we revealed the explicit sharp boundary, this result represents progress beyond the group of works mentioned at the end of \cref{sec:previous_research:positive_solution}.

Second, we propose a new procedure for proving positivity of third-order holonomic sequences. 
This procedure is applicable to a class of recurrences with a multiple dominant root.
Under a coefficient condition, we prove that the procedure halts on all initial values outside an exceptional plane (\cref{thm:procedure}).
This result partially extends \cref{thm:simple_unique_procedure}.

To prove the sufficiency (\cref{thm:positive_solution}) and to construct the procedure, 
we use the \emph{difference scaled by $S$} ($f(n+1)-S(n)f(n)$), where $S \in \R(x)$ (see \cref{sec:proof_strategy} for details).
To prove the necessity (\cref{thm:NonPositive}), we construct a third-order sequence without positivity as a linear combination of three first-order sequences.

\section{Preliminaries}
\label{sec:preliminaries}

\subsection{Orders based on asymptotic behavior}

Define an order $\geq$ on $\R(x)$ as $P(x)\geq Q(x)$ (or $P\geq Q$ for short) $\iff$ $P(x)\geq Q(x)$ for all sufficiently large $x$.
Define $\leq$ in a similar way.
We use $>$ and $<$ when we do not allow $P=Q$.

Let $d \in \Z$.
Define a quasi-order $ \succeq_d$ on $\R(x)$
as 
$P(x) \succeq_d Q(x)$ (or $P \succeq_d Q$ for short) $\iff$
$P(x) > Q(x) - Mx^{-d}$ for some $M > 0$.
Define $\preceq_d$ in a similar way.
We use $\succ_d$ and $\prec_d$ when we do not allow $P(x) = Q(x) +O(x^{-d})$, i.e., $\lvert P(x) - Q(x) \rvert = O(x^{-d}) \ (x \to \infty)$.
$P \succ_d Q$ if and only if, for any $M \in \R$, $P(x) > Q(x) + Mx^{-d}$.

\subsection{
Poincarization and normalization}
\label{sec:transformation}

The Positivity Problem for a solution $f$ of the recurrence~\eqref{eq:holonomic} is reduced to the one for $g(n) := f(n) \prod_{k=-r+1}^{n-r} R(k)^{-1}$, where $R \in \R(x)$ satisfies $R(n) > 0$ for all $n \geq -r + 1$, as follows.
(For $n=0$, we define $\prod_{k=-r+1}^{-r} R(k)^{-1} = 1$.)
$g$ satisfies a new recurrence
\[
g(n+r) = \frac{P_{r-1}(n)}{R(n)} g(n+r-1) + \dots + \frac{P_0(n)}{R(n)R(n-1) \dotsm R(n-r+1)} g(n),
\]
which is of Poincar\'{e} type if 
$\deg R = \lceil \max \left\{ \deg P_{r-1}, \frac12 \deg P_{r-2}, \dots, \frac1r \deg P_0 \right\} \rceil$.
Thus, 
we can assume that the recurrence is of Poincar\'{e} type in
\cref{thm:simple_unique_procedure,thm:simple_unique_posi_sol} without loss of generality. 
However, 
when ``$\max$'' inside the ceiling function is not an integer, 
the characteristic polynomial is $x^r$, whose dominant root is $0$ with multiplicity $r$, and thus this case does not fall within the scope of the theorems.

We consider only the recurrence~\eqref{eq:holonomic} with $r=3$ and $P_2 > 0$.
The Positivity Problem for the solutions of such recurrences is reduced to the one for the solutions of the normalized recurrence
\begin{equation}\label{eq:3rd-order}
	f(n+3) = f(n+2) + P(n) f(n+1) + Q(n) f(n),
\end{equation}
i.e., we can assume $P_2=1$ in the recurrence~\eqref{eq:holonomic}.
To see this, take $N \in \N$ such that $P_{2}(n+N) > 0$ for all $n \geq -2$, and reduce the Positivity Problem on $f$ to that of $\{ f(n+N) \}_{n \in \N}$.
Replace $n+N$ by $n$, and apply the above reduction with $R=P_2$. We obtain the above sequence $g$ which satisfies the recurrence~\eqref{eq:3rd-order}, where $P(x)=\frac{P_1(x)}{P_2(x)P_2(x-1)}$ and $Q(x)=\frac{P_0(x)}{P_2(x)P_2(x-1)P_2(x-2)}$.

Note that Poincarization and normalization are incompatible in general. 
We adopt normalization henceforth.

\subsection{Second-order holonomic recurrences}
\label{sec:preliminaries:2nd-order}

The following well-known theorem is a complete version of \cref{thm:simple_unique_posi_sol}
(see for instance \cite{Koo91,HK25} for the proof).
For simplicity, we state it only for normalized recurrences.

\begin{theorem} \label{thm:2nd_order}
	Let $P \in\R(x)$ satisfy $0 \notin P(\N)$. Then the recurrence
		\begin{equation} \label{eq:2nd-order_holonomic}
			f(n+2) = f(n+1) + P(n) f(n)
		\end{equation}
	admits a robustly ultimately positive solution if and only if
		\begin{equation} \label{eq:<=>exists_2nd-order_posi_sol}
			P(x) \succeq_3 -\frac14 - \frac{1}{16 x^2}.
		\end{equation}
\end{theorem}

In fact, this condition is equivalent to the existence of an ultimately positive solution (whose ultimate positivity is not necessarily robust) of the recurrence~\eqref{eq:2nd-order_holonomic}.
The threshold $- \frac14 - \frac1{16x^2}$ in the inequality~\eqref{eq:<=>exists_2nd-order_posi_sol} corresponds to the convergence condition for continued fractions~\cite[Theorem~2.8]{KKL+21}.
In \cref{thm:positive_solution}, we obtained a nontrivial threshold $R_P$ regarding the existence of a robustly ultimately positive solution of the third-order recurrences~\eqref{eq:3rd-order} for each $P$.

\section{Results}\label{sec:result}

\subsection{Robustly ultimately positive solutions}

\begin{theorem}\label{thm:positive_solution}
	Let $P, Q \in \R(x)$, and assume $0 \notin Q(\N)$. 
    If either of the following conditions holds, then the third-order recurrence~\eqref{eq:3rd-order}
    admits a robustly ultimately positive solution.
	\begin{enumerate}[(I) ]
		\item \label{item:P>=1} $1 \leq \lim\limits_{x \to \infty} P(x) \leq \infty$ and $Q \geq -P$.
		\item\label{item:P_converge} $-\frac13 - \frac1{9 x^2} \preceq_3 P(x) \prec_2 1 - \frac2x$ and $Q(x) \succeq_d R_P (x)$, where $R_P$ and $d$ are defined as follows.
        Let 
		\[
			P(x) = p_0 + \frac{p_1}x + \frac{p_2}{x^2} + \frac{p_3}{x^3} + O(x^{-4})
		\]
        be the Taylor expansion of $P(x)$ at $x=\infty$.
		\begin{enumerate}[(\ref{item:P_converge}-1) ]
			\item \label{item:-1/3<p0<=1} 
			If $-\frac13 < p_0$,
			\[
				R_P(x) = a^3 - a^2 - p_0 a  - \frac{p_1 a}x - \frac{b}{x^2},
				\quad\quad
				d=3,
			\]
			\[
				\text{where} \quad\quad
                a:=\frac{1+\sqrt{1+3p_0}}3, 
				\quad \quad
				b :=  p_2 a + \frac{(a+p_0+p_1)^2}{4 \sqrt{1+3p_0}}.
			\]
			\item \label{item:p0=-1/3_p1>0}  If $p_0=-\frac13$ and $p_1 > 0$,
			\[
			R_P(x) = \frac1{27}  -\frac{p_1}{3 x},
			\quad\quad
			d=2.
			\]
			\item \label{item:p0=-1/3_p1=0} If $(p_0, p_1) =\left(-\frac13, 0 \right)$,
			\[
			R_P(x) =  \frac1{27}  -\frac{p_2}{3 x^2} - \frac{c}{x^3},
			\quad\quad
			d=4,
			\quad\quad
			\text{where }\quad\quad
				c := \frac{p_2 + p_3}3 + \frac2{81 \sqrt{3}} (1+9p_2)^{3/2}. 
			\]
		\end{enumerate}
	\end{enumerate}
\end{theorem}

Under the condition $-\frac13 - \frac1{9 x^2} \preceq_3 P(x) \prec_2 1 - \frac2x $,
the scope of \cref{thm:simple_unique_posi_sol}, i.e., the condition equivalent to the uniqueness and simplicity of the dominant root, is $\lim_{x \to \infty} R_P(x) \allowbreak < \lim_{x \to \infty} Q(x) < \infty$.
(If the limits of $Q$ and $R_P$ coincide, there is a double or triple dominant root. If $\lim_{x \to \infty} Q(x) < \lim_{x \to \infty} R_P(x)$, 
there are two non-real complex dominant roots conjugate to each other.)
A slight extension following from the theorem in \cite{Ibr25} has the same scope under this condition.
Thus, our theorem covers additional cases.

In \eqref{item:-1/3<p0<=1}, 
$a$ is a root of the derivative of $\chi$, and so if $\lim\limits_{x \to \infty} Q(x) = \lim_{x \to \infty} R_P(x)$, $a$ is a double root of $\chi$ and is dominant.

\begin{example} \label{ex:2f1<f2}
Let 
\[
	f_1(n) := 2 n \sum_{k=0}^n \binom{n}{k}^3,
	\quad \quad
	f_2(n) := 8^n.
\]
Let us check that 
the (normalized version of the) recurrence of
$f(n) := f_2(n+4) - f_1(n+4)$ satisfies a condition in \cref{thm:positive_solution}.
In fact, our procedure proves that $f$ is a robustly positive solution (\cref{ex:auto-proof}).

Let $\mathbf{N}$ be a \emph{shift operator} such that $(\mathbf{N} g) (n) := g(n+1)$. 
\emph{Creative telescoping}~\cite{PWZ96} tells us that $f_1$ satisfies the recurrence
\[
	( \mathbf{N}^2 - U(n) \mathbf{N} - V(n) ) f_1 = 0,
	\quad\quad\quad
	U(x) := \frac{7 x^2 + 21 x + 16}{(x+1) (x+2)},
	\quad\quad
	V(x) := \frac{8(x+1)^2}{x (x+2)}
\]
for $n \geq 1$. $f_2$ satisfies $(\mathbf{N}-8) f_2 = 0$.
Then $f$ satisfies 
\[
	\bigg(   \mathbf{N}   - \underbrace{8 \cdot \frac{64 - 8 U(n+5) -  V(n+5)}{64 - 8 U(n+4) - V(n+4)}}_{\quad \quad \quad =: \ W(n) }   \bigg)   (\mathbf{N}^2 - U(n+4) \mathbf{N} - V(n+4)) f    =   0
\]
for $n \geq 0$.
Rewriting this in the form of \eqref{eq:general_3rd-order}, we obtain
\[
	P_2(x) = U(x+5) + W(x), 
	\quad\quad 
	P_1(x) = V(x+5) - W(x) U(x+4), 
	\quad\quad
	P_0(x) = - W(x) V(x+4).
\]
One can check $P_2(n)>0$ for all $n\geq -2$.
Next, we normalize this into the form of \eqref{eq:3rd-order} and obtain
\[
	P(x) = \frac{P_1(x)}{P_2(x) P_2(x-1)},
	\quad
	Q(x) = \frac{P_0(x)}{P_2(x) P_2(x-1) P_2(x-2)}.
\]
We can check $\infty \notin P(\N)$ and $0, \infty \notin Q(\N)$ step by step.
Since $\lim\limits_{x \to \infty} P(x) = -\frac{16}{75}$, we get $- \frac13 < \lim\limits_{x \to \infty} P(x) < 1$.
For the final step, we calculate with a computer to obtain
\[
	Q(x) = - \frac{64}{3375} - \frac{128}{5625 x} + \frac{19072}{253125 x^2} + O(x^{-3}),
	\quad\quad
	R_P(x)
	=
	- \frac{64}{3375} - \frac{128}{5625 x} + \frac{8272}{253125 x^2}.
\]

Note that $R_P(x)$ is the boundary in \cref{thm:positive_solution}~\eqref{item:-1/3<p0<=1}.
Thus $Q(x) \succeq_3 R_P(x)$, and so we can apply \cref{thm:positive_solution} to this recurrence.
On the other hand, since the constant terms of $Q(x)$ and $R_P(x)$ coincide, we cannot apply \cref{thm:simple_unique_posi_sol} or \ref{thm:simple_unique_procedure}.
\end{example}

Under the condition $\lim\limits_{x \to \infty} P(x) < 1$, we conjecture that the converse of \cref{thm:positive_solution} holds:

\begin{conjecture}\label{conj:}
    Let $P, Q \in \R(x)$ satisfy $\lim\limits_{x \to \infty} P(x) < 1$ and $0 \notin Q(\N)$. Then, 
    the recurrence~\eqref{eq:3rd-order} admits a robustly ultimately positive solution if and only if the condition~\eqref{item:P_converge} in \cref{thm:positive_solution} holds.
\end{conjecture}

The following theorems provide partial evidence for this.

\begin{theorem} \label{thm:NonPositive}
Let $P \in \R(x)$ satisfy $P \prec_3 0$ and the inequality~\eqref{eq:<=>exists_2nd-order_posi_sol}.
Let $R_P$ be as in \cref{thm:positive_solution}~\eqref{item:-1/3<p0<=1}.
Let $Q \in \R(x)$ satisfy $0 \notin Q(\N)$.
Then, the recurrence~\eqref{eq:3rd-order} admits a robustly ultimately positive solution if and only if $Q(x) \succeq_3 R_P(x)$.
\end{theorem}

The next theorem can be applied to a wider class of recurrences, but the conclusion is weaker than the above one.

\begin{theorem} \label{thm:Optimity}
For any $\varepsilon > 0$, $p_0 > -\frac13$ and $p_1, p_2 \in \R$, there exist $P,Q \in \R(x)$ with $P(x) = p_0 + \frac{p_1}x + \frac{p_2}{x^2} + O(x^{-3})$ and $Q(x) \geq R_P(x) - \frac{\varepsilon}{x^2}$ such that the recurrence~\eqref{eq:3rd-order} does not admit a robustly ultimately positive solution, where $R_P$ is as in \cref{thm:positive_solution}~\eqref{item:-1/3<p0<=1}.
\end{theorem}

Unfortunately, we cannot prove the necessity in cases \eqref{item:p0=-1/3_p1>0} and \eqref{item:p0=-1/3_p1=0} in \cref{thm:positive_solution}.

\subsection{A procedure for proving positivity}

The following theorem guarantees the termination of our procedure for all initial values outside an exceptional plane under a coefficient condition.

\begin{theorem}\label{thm:procedure}
	There exists a procedure for proving positivity of 
    third-order holonomic sequences which takes as input coefficients $(P,Q) \in (\overline{\Q} \cap \R)(x)^2$ of normalized recurrences satisfying the condition below
    and any initial values $(f(0), f(1), f(2)) \in  (\overline{\Q} \cap \R)^3$, and has the characterization of termination below.
	\begin{description}
		\item[Coefficient condition]
		$0 \notin Q(\N)$ and either of the following holds.
		\begin{enumerate}[(1)]
			\item \label{item:-1/4<=p0<=0} 
			The inequality~\eqref{eq:<=>exists_2nd-order_posi_sol} and $P \leq 0 < Q$ hold.			
			\item \label{item:-1/3<=p0<=-1/4}
			The inequality~\eqref{eq:<=>exists_2nd-order_posi_sol} does not hold, $P(x) \succ_2 -\frac13$ holds, and
            $Q(x) \succeq_d R_P(x)$ holds for $R_P$ and $d$ in \cref{thm:positive_solution}
            (in this case, the ones in \eqref{item:-1/3<p0<=1} or \eqref{item:p0=-1/3_p1>0}).
		\end{enumerate}
		\item[A characterization of termination]
		The procedure halts if and only if $(f(0), f(1), f(2)) \notin H(P, Q)$, or the holonomic sequence defined by the input is not positive, where $H(P, Q) \subseteq \R^3$ is a subspace with dimension $\leq 2$ depending on $P$ and $Q$.
	\end{description}
\end{theorem}

One can easily decide whether given $P$ and $Q$ satisfy each condition in the above theorem.
A sample illustration of the set of the initial values on which the procedure does not halt
is the right picture of \cref{fig:}. (Replace $H(P_0, P_1, P_2)$ in the picture with $H(P, Q)$.)
Like \cref{thm:positive_solution},
this theorem has a wider scope than \cref{thm:simple_unique_procedure} and the previous theorems in \cite{Ibr25}.
If we assume \cref{conj:}, the condition~\eqref{item:-1/3<=p0<=-1/4} is relaxed to the condition ``the inequality~\eqref{eq:<=>exists_2nd-order_posi_sol} does not hold and $P(x) \succ_2 -\frac13$ holds'', which does not depend on $Q$.

\begin{example} \label{ex:}
    How can we decide the positivity of
	\[
		f(n) := \sum_{k=0}^{\lfloor n/2 \rfloor} t^k \binom{2n-k}{n+k} \binom{n}{k}
	\]
    for a given $t \in \overline{\Q} \cap \R$ such that $-1 < t < 0$?
	Creative telescoping tells us that $f$ satisfies a recurrence of the form~\eqref{eq:general_3rd-order}, and the coefficients $P_2(x), P_1(x), P_0(x) \in \R(x)$ converge to the following values as $x \to \infty$, respectively.
	\[
		-\frac{t^2-22t+1}{4t},
		\quad\quad
		\frac{11 t^2 - 11 t + 2}{2 t},
		\quad\quad
		\frac{(4 t - 1) t}{4}.
	\]
    Since $-\frac{t^2-22t+1}{4t} > 0$, we can normalize this recurrence to the one of the form~\eqref{eq:3rd-order} as in \cref{sec:transformation} (after shifting $n$ suitably).
    The coefficients $P$ and $Q$ converge to 
    \[
		p_0 := \frac{8(11 t^2 - 11 t + 2) t}{(t^2-22t+1)^2},
		\quad\quad
		q_0 := -\frac{16 (4 t - 1) t^4}{(t^2-22t+1)^3},
	\]
    respectively.
    By $-1 < t < 0$, we have $-\frac13 < p_0 < 0$.
    Let $\alpha$ be the unique $t \in (-1, 0)$ such that 
	$t^3 - 24 t^2 + 3 t + 1 = 0$.
    Then, for $R_P$ of \cref{thm:positive_solution}~\eqref{item:-1/3<p0<=1},
    $q_0 > \lim\limits_{x \to \infty} R_P(x)$ if $t>\alpha$, $q_0 = \lim\limits_{x \to \infty} R_P(x)$ if $t = \alpha$, and $q_0 < \lim\limits_{x \to \infty} R_P(x)$ if $t<\alpha$.
    In the case $t>\alpha$, by \cref{thm:simple_unique_procedure},
    we can decide the positivity of $f$ (as long as the initial value is outside the exceptional plane).
    In the case $t<\alpha$,
    by \cref{conj:},
    we expect that $f$ is not positive (as long as the initial value is outside a set contained in a plane), and so it is a good strategy to seek $n$ such that $f(n) \leq 0$.
    
    The case $t=\alpha$ is where \cref{thm:procedure} comes into play.
    Using a computer, we can verify $-\frac13 < p_0 < -\frac14$ and $Q(x) \succeq_3 R_P (x)$.
    Thus $P$ and $Q$ satisfy the condition~\eqref{item:-1/3<=p0<=-1/4} in \cref{thm:procedure}.
    See \cref{sec:SpecificCoefficients} for the values and the Taylor expansions of $P_2$, $P_1$ and $P_0$, the Taylor expansions of $P$ and $Q$, and $R_P$.

    Note that, in the case $t = \alpha$, we do not have to shift $n$ for the normalization since $P_2(n) > 0$ for all $n \geq -2$.
    Our procedure succeeded in proving the positivity of $f$ (see \cref{ex:auto-proof2}).
\end{example}

\section{Method}
\label{sec:proof_strategy}

In this section, we overview the method used to prove \cref{thm:positive_solution,thm:procedure} and compare it with previous research.
Most of the previous research in 
\cref{sec:positivity_proving_procedure} used the \emph{positivity certificate} $\cert(n)$.
This is a condition of the form ``$f(n)>0$ $\land$ \dots'' with $N \in \N$
that satisfies $\cert(n) \implies \cert(n+1)$ for all $n \geq N$.
Once $f$ satisfies $\cert(n)$ for some $n \geq N$, then $f(n) > 0$ is guaranteed thereafter.
This gives the existence of an ultimately positive solution if, for example, we can choose the initial values of $f$ such that $f$ satisfies $\cert(N)$.
Moreover, it leads to a procedure for proving positivity below (\cref{proc:overview}).

\begin{algorithm}[htb]
\caption{Proving positivity}
\label{proc:overview}
\begin{algorithmic}[1] 
\AlgInput{$(P_0, \dots, P_{r-1}) \in (\overline{\Q} \cap \R)(x)^r$ and $(f(0), \dots , f(r-1)) \in (\overline{\Q} \cap \R)^r$.}
\AlgOutput{Whether $f$ defined by the input is positive.}

	\State{Find $N$ such that $\cert(n) \implies \cert(n+1)$ for all $n \geq N$.}\label{state:findN}
	
	\State{$n \leftarrow 0$}
		
	\While{\label{state:while_loop} $f(n) > 0 \ \text{and} \ ( n < N   \ \text{or} \ \lnot \cert(n) )$}
		
		\State $n \leftarrow n+1$
	
	\EndWhile
		
	\State If $f(n) \leq 0$, output ``No''. If $\cert(n)$ and $n \geq N$, output ``Yes''.
\end{algorithmic}
\end{algorithm}

Positivity certificates for general-order holonomic sequences 
have so far been constructed using quantifier elimination~\cite{GK05,KP10,Pil13}
and contracted cones~\cite{IS26,Ibr25}.
For second-order sequences, a certificate of the form $F(n) > 0 \land \Delta_L F(n) > 0$ for some function $L$ is used implicitly \cite{HK25,KKL+21}, where $\Delta_L F$ is defined below.
While $L$ is not a rational function in the literature, we will find that we can choose $L$ as a rational function.

\begin{definition} \label{def:S-deference}
	Let $F$ be a real-valued sequence and $L \in \R(x)$. The {\boldmath \textbf{$L$-scaled difference}} $\Delta_L F$ of $F$ is the sequence defined as $\Delta_L F(n) := F(n+1) - L(n) F(n)$.
\end{definition}

In this paper, we construct a certificate of the form 
\[
f(n) > 0 \land \text{``$F = \Delta_S f$ satisfies the certificate for second-order sequences''},
\] 
that is,
	\begin{equation} \label{eq:positivity_certificate}
		\cert(n) := \quad \quad
		f(n) > 0 \ \land \ \Delta_S f(n) > 0 \ \land \ \Delta_L \Delta_S f(n) > 0.
	\end{equation}
$S$ and $L$ are chosen so that $\cert(n) \implies \cert(n+1)$ and other conditions we will use hold.

Note that we omitted some minor tasks and specifications as follows. 
$N$ should be chosen so that $S$ and $L$ do not have poles in $\Z \cap [N, \infty)$, and the procedure does not evaluate $\cert(n)$ when $n<N$. 
If necessary, we assume that the procedure takes as input a coefficient list of a normalized recurrence.
The requirement on $S$ and $L$ is specified in the proof of \cref{lem:sufficient_condition}. 
See also \cref{sec:execution} for sample executions involving these minor notes. 

\section{Proof}

\subsection{Proof of Theorems~\ref{thm:positive_solution} and \ref{thm:procedure}}

In \cref{sec:proof_of_thm:positive_solution}, we will show  \cref{thm:positive_solution,thm:procedure}.
Therein, we use \cref{lem:sufficient_condition,lem:algorithmSL}, which state that there exist $S$ and $L$ such that the certificate~\eqref{eq:positivity_certificate} satisfies suitable conditions. 
In
\cref{sec:proof_of_main_lemmas},
we will show
\cref{lem:sufficient_condition,lem:algorithmSL}. 
Choosing $S$, $L$ satisfying an appropriate inequality leads to the suitable certificate. 
\cref{lem:2nd-order,lem:QE,lem:pre-algorithmSL} state the existence of such $S$, $L$, and they are proved in \cref{sec:proofs_of_lemmas}.

\subsubsection{Proof of Theorems~\ref{thm:positive_solution} and \ref{thm:procedure} with the positivity certificate}
\label{sec:proof_of_thm:positive_solution}

The next lemma states the existence of the positivity certificate. We will use it to prove \cref{thm:positive_solution}.

\begin{lemma}
\label{lem:sufficient_condition}
	Let $P, Q \in \R(x)$ satisfy $0 \notin Q(\N)$ and 
    condition~\eqref{item:P_converge} in \cref{thm:positive_solution}.
    Then there exist $S, L \in \R(x)$ and $N \in \N$ such that, for all real-valued $f$ satisfying the recurrence~\eqref{eq:3rd-order}, the certificate~\eqref{eq:positivity_certificate} satisfies $\cert(n) \implies \cert(n+1)$ for all $n \geq N$.
\end{lemma}

\begin{proof}[Proof of \cref{thm:positive_solution}]
Let $P$ and $Q$ be as in the statement.
If they satisfy the condition~\eqref{item:P>=1}, transform the recurrence~\eqref{eq:3rd-order} into
\[
	f(n+3) - f(n+2) = P(n) (f(n+1) - f(n)) + (P(n)+Q(n)) f(n).
\]
There exists $N \in \N$ such that $P(n) > 0$ and $P(n) + Q(n) \geq 0$ for all $n \geq N$.
If we choose a suitable $f$ so that $0 < f(N) < f(N+1) < f(N+2)$, then we get $0<f(n) < f(n+1) < f(n+2)$ inductively. 
Hence $f$ is a robustly ultimately positive solution.

If $P$ and $Q$ satisfy the condition~\eqref{item:P_converge}, 
take $S$, $L$, $N$ as in 
\cref{lem:sufficient_condition}.
We can choose a solution $f$ of the recurrence~\eqref{eq:3rd-order} such that the certificate~\eqref{eq:positivity_certificate} holds for $n = N$ since $0 \notin Q(\N)$.
Then $f$ is a robustly ultimately positive solution.
\end{proof}

The following lemma states that a condition concerning the certificate and one concerning the growth rate of $f$ hold simultaneously.

\begin{lemma} \label{lem:algorithmSL}
	There is an algorithm that,  given $P, Q \in (\overline{\Q} \cap \R) (x)$ satisfying the coefficient condition of \cref{thm:procedure} and $P \neq 0$, computes  
    $S, L \in (\overline{\Q} \cap \R) (x)$ and $N \in \N$ that satisfy not only the property in \cref{lem:sufficient_condition}
    but also the following.
    For any positive solution $\{ f(n) \}_{n \geq N}$ of the recurrence~\eqref{eq:3rd-order}, 
    there exists $n \geq N$ such that the certificate~\eqref{eq:positivity_certificate} holds 
    if and only if $\widetilde{f}_{S,L}$ is unbounded, where
	\begin{equation}
	\label{eq:adjusted_f}
	\widetilde{f}_{S,L}(n) =
	\begin{cases}
	f(n) \prod_{k=N}^{n-1} S(k)^{-1} & \text{if $S \neq 0$}
	\\
	f(n) \prod_{k=N}^{n-2} L(k)^{-1} & \text{if $S = 0$}
	\end{cases} \ .
	\end{equation}
\end{lemma}

\begin{proof}[Proof of \cref{thm:procedure}]
Let us first consider the case of $P=0$. Set $H(0, Q) := \{ (0, 0, 0) \}$ and
\[
	\cert(n) := \quad \quad f(n) > 0 \ \land \ f(n+1) > 0 \ \land \ f(n+2) > 0.
\]
Then \cref{proc:overview} decides the positivity. Here, $N$ is a natural number such that $Q(n) \geq 0$ for all $n \geq N$. 
We assume $P \neq 0$ from now on.

Compute $S$, $L$, $N$ with the algorithm in \cref{lem:algorithmSL} with input $P$, $Q$. Set
\[
H(P, Q) = \left\{ (f(0), f(1), f(2)) \in \R^3 \ \middle| \ \text{$\widetilde{f}_{S, L}$ defined as in \eqref{eq:adjusted_f} is bounded} \right\},
\]
where $f$ in the definition~\eqref{eq:adjusted_f} represents the third-order sequence defined by the initial value $(f(0), f(1), f(2))$ and the recurrence~\eqref{eq:3rd-order}.
$H(P, Q)$ is a subspace of $\R^3$. 
Moreover, $H(P, Q)$ has dimension $\leq 2$, since if we choose $f$ such that the certificate~\eqref{eq:positivity_certificate} is satisfied for $n = N$, then $\widetilde{f}_{S, L}$ is unbounded, and so $H(P, Q) \neq \R^3$.

As a procedure in the statement, we propose \cref{proc:overview} with the certificate~\eqref{eq:positivity_certificate}.
When it halts, the output is obviously correct.
Let us confirm the characterization of termination.
It does not terminate if and only if $f$ is positive and there is no $n \geq N$ such that the certificate~\eqref{eq:positivity_certificate} holds.
By \cref{lem:algorithmSL}, this is equivalent to the positivity of $f$ and the boundedness of $\widetilde{f}_{S, L}$.
This means that $f$ is positive and $(f(0), f(1), f(2)) \in H(P, Q)$, which is the negation of the desired condition.
\end{proof}

\subsubsection{
Inequalities that lead to the positivity certificates}
\label{sec:proof_of_main_lemmas}

We will show \cref{lem:sufficient_condition,lem:algorithmSL} by using \cref{lem:2nd-order,lem:QE,lem:pre-algorithmSL}.
These lemmas state the existence of $S$, $L$ that satisfy certain inequalities.
In the literature, \cref{lem:2nd-order} is used for proving positivity of second-order sequences and to prove \cref{thm:2nd_order}.

\begin{lemma}\label{lem:2nd-order}
Let $A, B \in \R(x)$ and assume $B > 0$. $P(x) = \frac{A(x)}{B(x)B(x-1)}$ satisfies the inequality~\eqref{eq:<=>exists_2nd-order_posi_sol} if and only if there exists $L \in \R(x)$ satisfying the following two conditions.
\begin{enumerate}[(1)]
	\item \label{eq:0<L<=B} $0 < L(x+1) < B(x)$
	\item \label{eq:L(1-L)+A>=0} $L(x)(B(x)-L(x+1)) + A(x) \geq 0$
\end{enumerate}
\end{lemma}

The symbols $A$, $B$ come from the \emph{canonical numerators} and \emph{canonical denominators} of continued fractions~\cite{LW08}.

\begin{lemma} \label{lem:QE}
	Assume that $P, Q \in \R(x)$ satisfy $0 \notin Q(\N)$ and the condition~\eqref{item:P_converge} in \cref{thm:positive_solution}.
    Then there exists $S \in \R(x)$ satisfying the following three conditions.
	\begin{enumerate}[(1)]
		\item \label{eq:0<=S<1}
		$S(x) \geq 0$
		\item \label{eq:A/BB>=-1/4}
		$B(x) := 1-S(x+2)$ and $A(x):=B(x) S(x+1) + P(x)$ satisfy one (i.e., both) of the equivalent conditions in \cref{lem:2nd-order}.
		\item \label{eq:C>=0}
		$C(x) := A(x) S(x) + Q(x) \geq 0$
	\end{enumerate}
\end{lemma}

\begin{proof}[Proof of \cref{lem:sufficient_condition}]
For $P$, $Q$ in the statement, take $S$, $A$, $B$, $C$ as in \cref{lem:QE}.
Take $N' \in \N$ such that the conditions~\eqref{eq:0<=S<1},~\eqref{eq:C>=0} in the lemma literally hold for all $x \geq N'$.
Take $L$ as in \cref{lem:2nd-order}, and again, take $N'' \in \N$ such that the conditions~\eqref{eq:0<L<=B},~\eqref{eq:L(1-L)+A>=0} in the lemma literally hold for all $x \geq N''$.
Set $N := \max\{ N', N''+1 \}$.

Assume $\cert(n), \ n \geq N$.
Rewrite the recurrence~\eqref{eq:3rd-order} as
\begin{equation}
\label{eq:3rd->1st_and_2nd}
\begin{aligned}
	\Delta_L \Delta_S f(n+1) = (B(n) &- L(n+1)) \Delta_L \Delta_S f(n) 
	\\
	&+ (L(n)(B(n)-L(n+1)) + A(n))\Delta_S f(n) + C(n) f(n).
\end{aligned}
\end{equation}
Since $B(n) - L(n+1) > 0$, $L(n)(B(n)-L(n+1)) + A(n) \geq 0$ and $C(n) \geq 0$, it follows that $\Delta_L \Delta_S f(n+1) > 0$. Since $\Delta_L \Delta_S f(n) > 0 $ and $\Delta_S f(n) > 0$, it follows that $\Delta_S f(n+1) > 0$. Since $\Delta_S f(n) > 0 $ and $ f(n) > 0$, it follows that $f(n+1) > 0$. Thus we obtain $\cert(n+1)$.
\end{proof}

The idea used in the proof of \cref{lem:algorithmSL} is basically the same as that of \cref{lem:sufficient_condition}, except that we need some additional conditions.

\begin{lemma} \label{lem:pre-algorithmSL}
	There exists an algorithm that, given $P, Q \in (\overline{\Q} \cap \R) (x)$ satisfying $P \neq 0$ and the coefficient condition of    
    \cref{thm:procedure}, 
    outputs     
    $S, L \in (\overline{\Q} \cap \R) (x)$
    that satisfy not only the conditions~\eqref{eq:0<=S<1},~\eqref{eq:C>=0} in \cref{lem:QE} and the two conditions in \cref{lem:2nd-order} but also $A \leq 0$ and the following inequalities.
    Here, we define $A$ and $B$ as in the condition~\eqref{eq:A/BB>=-1/4} in \cref{lem:QE}.
	\begin{align} 
		\label{eq:L/S<1-1/x<(1-S-L)/S}
		\frac{L(x)}{S(x+1)} \prec_2  1 - \frac1x  &\preceq_2 \frac{B(x) - L(x+1)}{S(x+2)} & \text{ if $S \neq 0$}
		\\
		\label{eq:L<1-1/x<B/L-1}
		1 - \frac1x &\preceq_2 \frac{B(x) - L(x+1)}{L(x+1)} 
		& \text{if $S = 0$}
	\end{align}
\end{lemma}

In the following proof, we use the fact that, for any $\alpha \in \R$ and any function $U \colon \N \to \R$ with $\lvert U(l) \rvert = O(l^{-2})$ and $1 + \frac{\alpha}l + U(l) \neq 0$ for all $l \in \N_{>0}$,
\[
	\sum_{k=1}^{n-1}\prod_{l=1}^{k-1} \left( 1 + \frac{\alpha}l + U(l) \right) = O(1) \iff \alpha < -1.
\]
One can verify this by $\prod_{l=1}^{k-1} \lvert 1 + \frac{\alpha}l + U(l) \rvert= \Theta \left( k^{\alpha}\right)$ and $\sum_{k=1}^{\infty} k^{\alpha} < \infty$ $\iff$ $\alpha < -1$.
The threshold $1 - \frac1x$ in the above lemma comes from this.

\begin{proof}[Proof of \cref{lem:algorithmSL}]
Given $P$ and $Q$, use the algorithm in \cref{lem:pre-algorithmSL} to compute $S$, $L$.
Define $A$, $B$ and $C$ as in \cref{lem:QE}.
Compute $N \in \N$ such that the conditions \eqref{eq:0<=S<1} and \eqref{eq:C>=0} in \cref{lem:QE} and the two conditions in \cref{lem:2nd-order}
literally hold for all $x \geq N - 1$.
If $S\neq0$, use a larger $N$ such that $S(n) > 0$ for all $n \geq N$. 
The discussion in the proof of 
\cref{lem:sufficient_condition}
is valid now, so $S$ and $L$ satisfy the property stated in \cref{lem:sufficient_condition}.

Now, it remains to show the equivalence in the statement for all positive solutions $\{ f(n) \}_{n \geq N}$ of the recurrence~\eqref{eq:3rd-order}.
First, assume that there exists $n \geq N$ such that $\cert(n)$ holds.
Replacing $N$ by a larger integer, we can assume $\cert(N)$.
Replacing $n$ with $l$ in \cref{eq:3rd->1st_and_2nd}, we obtain $\Delta_L \Delta_S f(l+1) \geq (B(l) - L(l+1)) \Delta_L \Delta_S f(l)$.
Hence $\Delta_L \Delta_S f(k) \geq \big( \Delta_L \Delta_S f(N)  \big)\prod_{l=N}^{k-1} (B(l) - L(l+1))$
for all $k \geq N$.
Multiply both sides by $\prod_{l=N}^{k} L(l)^{-1}$ and get
\begin{equation}
\label{eq:estimate_Delta_S_f}
	\big( \Delta_S f(n) \big)   \prod_{l=N}^{n-1} L(l)^{-1} 
	= 
	\Omega \left(  
	\sum_{k=N}^{n-1} \prod_{l=N}^{k-1} \frac{B(l) - L(l+1)}{L(l+1)}
	\right).
\end{equation}
If $S=0$, the left-hand side is $\widetilde{f}_{S,L}(n+1)$ and the function inside $\Omega$ on the right-hand side diverges to $\infty$ as $n \to \infty$ due to the inequality~\eqref{eq:L<1-1/x<B/L-1}.
If $S \neq 0$, in \cref{eq:estimate_Delta_S_f}, retain only the term for $k=n-1$ in the summation and simplify it. Then we get $\Delta_S f(n) = \Omega\left(  \prod_{l=N}^{n-2}(B(l) - L(l+1)) \right) $.
A similar calculation to the above gives us
\[
	\widetilde{f}_{S,L}(n) = f(n)  \prod_{k=N}^{n-1} S(k)^{-1} 
	=
	\Omega \left( 
	\sum_{k=N}^{n-1} \prod_{l=N}^{k-2} \frac{B(l) - L(l+1)}{S(l+2)}
	\right).
\]
It follows from the inequality~\eqref{eq:L/S<1-1/x<(1-S-L)/S} that $\widetilde{f}_{S,L}(n)  \to \infty$.

Next, let us show the inverse. Assume that there is no $n \geq N$ such that $\cert(n)$ holds.
Then, for all $n \geq N$, we have $\Delta_S f(n) \leq 0$ or $\Delta_L \Delta_S f(n) \leq 0$.

Since $A(n) \leq 0 \leq C(n)$ for all sufficiently large $n$, once we have $\Delta_S f(n) \leq 0 < \Delta_S f (n+1)$, we also have $\Delta_L \Delta_S f(n+1) >0$ as shown below.
\begin{align*}
	\Delta_S f(n+2) 
	&=
	B(n) \Delta_S f(n+1) + A(n) \Delta_S f(n) + C(n) f(n)
	\\
	&\geq 
	B(n) \Delta_S f(n+1) 
	>
	L(n+1) \Delta_S f(n+1)
\end{align*}
In this calculation, of course, we used the recurrence~\eqref{eq:3rd-order}.
Thus we have $\Delta_S f(n) \leq 0$ for all sufficiently large $n$, or $\Delta_S f(n) > 0$ and $\Delta_L \Delta_S f(n) \leq 0$ for all sufficiently large $n$.
In the former case, if we assume $S=0$, then $f(n+1) \leq 0$. This is a contradiction, so $S \neq 0$.
Since $f(n) \prod_{k=N}^{n-1} S(k)^{-1}$ is non-increasing and positive, it is bounded.
In the latter case, a similar calculation shows that $\Delta_S f(n) \prod_{k=N}^{n-1} L(k)^{-1}$ is non-increasing and positive, and so bounded.
Then, if $S=0$, the sequence $f(n+1) \prod_{k=N}^{n-1} L(k)^{-1}$ is bounded. If $S \neq 0$, we have $\Delta_S f(n) = O \left( \prod_{k=N}^{n-1} L(k) \right)$, which means
\[
	f(n) \prod_{k=N}^{n-1} S(k)^{-1} = O \left(  \sum_{k=N}^{n-1} \prod_{l=N}^{k-1} \frac{L(l)}{S(l+1)}  \right).
\]
It follows from the inequality~\eqref{eq:L/S<1-1/x<(1-S-L)/S} that $f(n) \prod_{k=N}^{n-1} S(k)^{-1} = O(1)$.
\end{proof}

\subsubsection{Existence of rational functions satisfying certain inequalities} \label{sec:proofs_of_lemmas}

We will show \cref{lem:2nd-order,lem:QE,lem:pre-algorithmSL} by constructing $S$ and $L$ explicitly.

\begin{proof}[Proof of \cref{lem:2nd-order}]
    It suffices to show this lemma only for $B=1$ because, for general $B$, we can reduce the lemma to the case with $B=1$ by replacing $A(x)$ by $\frac{A(x)}{B(x)B(x-1)}$ and $L(x)$ by $\frac{L(x)}{B(x-1)}$.

($\implies$) Let $p_3$ be the coefficient of $x^{-3}$ in the Laurent expansion of $P(x) = A(x)$ at $x=\infty$. Set
\begin{equation} \label{eq:L=1/2+1/4x+}
	L(x) := \frac12 + \frac1{4x} + \frac{-2p_3 + 1/4}{x^2}.
\end{equation}
Obviously $0 < L < 1$. The other condition~\eqref{eq:L(1-L)+A>=0} follows from 
\[
	L(x)(1-L(x+1)) 
	=\frac14 + \frac1{16 x^2} + \frac{-p_3 + 1/16}{x^3} + O(x^{-4})
\]
and the inequality~\eqref{eq:<=>exists_2nd-order_posi_sol}.

($\impliedby$) Since $0 < L < 1$, there exists $\alpha := \lim\limits_{x \to \infty} L(x) \in \R$. 
Letting $x \to \infty$ in the inequality~\eqref{eq:L(1-L)+A>=0} gives us $\alpha(1 - \alpha) + \lim\limits_{x \to \infty} A(x) \geq 0$.
Since $\alpha(1 - \alpha) \leq \frac14$,
it follows that $\lim\limits_{x \to \infty} P(x) = \lim\limits_{x \to \infty} A(x) \geq -\frac14$. If $\lim\limits_{x \to \infty} P(x) > -\frac14$, the proof is done. Assume $\lim\limits_{x \to \infty} P(x) = -\frac14$. Then we have $\alpha = \frac12$. 
Expanding $L(x) = \frac12 + \frac{l_1}x +  O(x^{-2})$, we get
\[
	L(x) (1 - L(x+1)) = \frac14 + \frac{l_1 / 2 - l_1^2}{x^2} + O(x^{-3}).
\]
The inequality~\eqref{eq:<=>exists_2nd-order_posi_sol} follows from this and $l_1 / 2 - l_1^2 \leq \frac1{16}$.
\end{proof}

\begin{proof}[Proof of \cref{lem:QE}]
For each case in \cref{thm:positive_solution}~\eqref{item:P_converge}, we construct $S \in \R(x)$ satisfying the three conditions.

\eqref{item:-1/3<p0<=1} 
Set
\begin{equation} \label{eq:construction_of_S}
	S(x) = a + \frac{s_1}{x} + \frac{s_2}{x^2}, 
	\quad \quad 
	s_1 := \frac{a + p_0 + p_1}{2 \sqrt{1 + 3 p_0}}.
\end{equation}
We will choose $s_2$ later. 
$S\geq 0$ follows from $a > 0$.
Let us next check the condition~\eqref{eq:A/BB>=-1/4}.
By $p_0 + \frac{p_1}x < 1 - \frac2x$, we have $a + \frac{s_1}x < 1$, and so $B > 0$.
Let us show that $\frac{A(x)}{B(x)B(x-1)}$ satisfies the inequality~\eqref{eq:<=>exists_2nd-order_posi_sol}.
If $p_0= 1$, by $S(x) \to 1$, we have $A(x) \to 1$ and so $ \frac{A(x)}{B(x)B(x-1)} >0$.
If $p_0 < 1$, by $\frac13 < a<1$ and $-3 a^2 + 2 a + p_0 = 0$, we have
\[
	\lim_{x \to \infty} \frac{A(x)}{B(x)B(x-1)} 
	=
	\frac{a(1-a) + p_0}{(1-a)^2}
	=
	\frac{2a^2 - a}{(1-a)^2}
	=
	-\frac14 + \frac14 \left( \frac{3a - 1}{1-a} \right)^2
	>
	-\frac14.
\]
Finally, we show the condition~\eqref{eq:C>=0}.
Expand $C(x)$ as follows:
\begin{align*}
	C(x) &= Q(x) -a^3 + a^2 + p_0 a + \frac{p_1 a}{x} + \frac{p_2 a + (a + p_0 + p_1) s_1 - \sqrt{1+3 p_0} s_1^2}{x^2} + O(x^{-3}) 
	\\
	&=
	Q(x) -a^3 + a^2 + p_0 a + \frac{p_1 a}{x} + \frac{b}{x^2} + O(x^{-3}) \succeq_3 0.
\end{align*}
We chose the value of $s_1$ that maximizes the coefficient of $x^{-2}$.
Since the coefficient of $s_2 / x^3$ in the expansion of $C(x)$ is $a \sqrt{1 + 3 p_0} > 0$, for a sufficiently large $s_2$, we have $C \geq 0$.

\eqref{item:p0=-1/3_p1>0}
Set
\[
	S(x) = \frac13 + \frac{s_1}x.
\]
We will choose $s_1$ later. This satisfies the condition~\eqref{eq:0<=S<1}.
The condition~\eqref{eq:A/BB>=-1/4} holds as
\[
	\frac{A(x)}{B(x)B(x-1)} 
	= 
	 -\frac14 + \frac94 \frac{p_1}{x} + O(x^{-2})
\]
independently of $s_1$.
The condition~\eqref{eq:C>=0} also holds as
\[
	C(x) = Q(x) -\frac1{27}+ \frac{p_1/3}x + \frac{p_2/3 + p_1 s_1}{x^2} + O(x^{-3})
\]
for a sufficiently large $s_1$.

\eqref{item:p0=-1/3_p1=0} 
Set
\[
	S(x) = \frac13 + \frac{s_1}{x} + \frac{s_2}{x^2},
	\quad\quad
	s_1 = \frac13 - \frac2{3 \sqrt3} \sqrt{1 + 9 p_2} \ .
\]
We choose the value of $s_2$ later.
This satisfies the condition~\eqref{eq:0<=S<1}. The condition~\eqref{eq:A/BB>=-1/4} follows from
\[
	\frac{A(x)}{B(x)B(x-1)} 
	= 
	 -\frac14 + \frac{9}{16} \frac{4 p_2 + 2 s_1 - 3 s_1^2 }{x^2} + O(x^{-3})
	=
	-\frac14 - \frac{1}{16 x^2} + O(x^{-3}).
\]
Let us consider the condition~\eqref{eq:C>=0}.
Expand $C(x)$ as
\begin{align*}
	C(x) 
	&= 
	Q(x) -\frac1{27} + \frac{p_2/3}{x^2} + \frac{\frac13 p_3 + \left( p_2 - \frac29 \right) s_1 + s_1^2 - s_1^3}{x^3}
	+O(x^{-4})
	\\
	&=
	Q(x) -\frac1{27} + \frac{p_2/3}{x^2} + \frac{c}{x^3}
	+O(x^{-4})
	\succeq_4 0.
\end{align*}
We chose the value of $s_1$
that maximizes the coefficient of $x^{-3}$ in the expansion above subject to the condition that the coefficient of $x^{-2}$ in the Taylor expansion of $\frac{A(x)}{B(x)B(x-1)}$ is $\geq -\frac1{16}$.
The coefficient of $s_2 / x^4$ in the expansion of $C(x)$ is
\[
	-\frac49 - 3 p_2 - \frac2{3 \sqrt3} \sqrt{1 + 9 p_2}
	\quad \left( \leq -\frac19 \right),
\]
so $C \geq 0$ for a sufficiently negative $s_2$.
\end{proof}

\begin{proof}[Proof of \cref{lem:pre-algorithmSL}]
For $P, Q \in (\overline{\Q} \cap \R) (x)$ satisfying $P \neq 0$ and the coefficient condition of \cref{thm:procedure}, we construct $S$, $L$ as follows.
We divide the construction based on which coefficient condition is satisfied by $P$ and $Q$.

If $P$ and $Q$ satisfy the coefficient condition~\eqref{item:-1/4<=p0<=0}, we define $S=0$ and $L$ as in the definition~\eqref{eq:L=1/2+1/4x+}.
Then it is obvious that the conditions~\eqref{eq:0<=S<1} and \eqref{eq:C>=0} are satisfied. Moreover, the two conditions in \cref{lem:2nd-order} follow from a similar discussion to that in its proof.
In addition, $A(x) = P(x) \leq 0$ holds.
It remains to show the inequality~\eqref{eq:L<1-1/x<B/L-1}.
Since $B=1$ now, we have
\[
	\frac{1 - L(x+1)}{L(x+1)} 
	=
	\frac{\frac12 - \frac1{4x}}{\frac12 + \frac1{4x}} + O(x^{-2})
	= 1 - \frac1x +O(x^{-2}).
\]

If $P$ and $Q$ satisfy the coefficient condition~\eqref{item:-1/3<=p0<=-1/4}, once we construct appropriate $S>0$ and $L$, the inequality $A \leq 0$ automatically follows from
\cref{lem:2nd-order} with $A(x)$, $B(x)$, $L(x)$ replaced by $P(x)$, $1$, $S(x+1)$, and 
the fact that 
$P$ does not satisfy the inequality~\eqref{eq:<=>exists_2nd-order_posi_sol}.

Define $S$ as in the proof of \cref{lem:QE}.
We already checked that conditions~\eqref{eq:0<=S<1} and \eqref{eq:C>=0} of
\cref{lem:QE} are satisfied.
We adopt different constructions of $L$ depending on which of \eqref{item:-1/3<p0<=1} and \eqref{item:p0=-1/3_p1>0} holds.

\eqref{item:-1/3<p0<=1}
Set
\[
	L(x) = \left( 1 - \frac{a}{1-a} +  \frac{l_1}x + \frac{l_2}{x^2} \right) B(x-1),
	\quad \quad
	l_1 = \frac{ (a + p_0) (1 - 2a) - p_1}{2 (1-a)^2 \sqrt{1+3p_0} }.
\]
We will choose $l_2$ later. 
Let us first show that the two conditions of \cref{lem:2nd-order} hold. Condition~\eqref{eq:0<L<=B}: Since $\frac13 < a \leq \frac12$, we have $L(x+1) < B(x)$. In addition, $0 \leq 1 - \frac{a}{1-a}$ holds.
If this is a strict inequality, $L>0$. If equality holds, then $p_0 = -\frac14$ and $a=\frac12$. Since the inequality~\eqref{eq:<=>exists_2nd-order_posi_sol} does not hold, $p_1 \leq 0$. If $p_1<0$, we have $l_1 = -4 p_1 > 0$ and $L(x) = \left( \frac{l_1}x + \frac{l_2}{x^2} \right) B(x-1)>0$.
If $p_1=0$, we have $L>0$ for all sufficiently large $l_2$.
Condition~\eqref{eq:L(1-L)+A>=0}: 
By a straightforward calculation, $L(x) (B(x)-L(x+1)) + A(x) = O(x^{-2})$.
Since the coefficient of $l_2 / x^2$ in the expansion of $L(x) (B(x)-L(x+1)) + A(x)$ is $(1-a)(3a-1) \ (> 0)$, again, any sufficiently large $l_2$ makes it positive.
The inequality~\eqref{eq:L/S<1-1/x<(1-S-L)/S} follows from:
\begin{align*}
	&\lim_{x \to \infty} \frac{L(x)}{S(x+1)} = \frac{1-a}a \left( 1 - \frac{a}{1-a} \right)
	< 2 \cdot \frac12 = 1,
	\\
	&\frac{B(x) - L(x+1)}{S(x+2)} 
	=
	1 - \frac{s_1 + (1-a)^2 l_1}{a(1-a)} \frac1x + O(x^{-2})
	=
	1- \frac1x + O(x^{-2}).
\end{align*}

\eqref{item:p0=-1/3_p1>0}
Set
\[
	L(x) = \left( \frac12 + \frac{l_1}x \right)  B(x-1).
\]
We will choose the value of $l_1$ later. 
Let us first show that the two conditions of \cref{lem:2nd-order} hold. Condition~\eqref{eq:0<L<=B} holds obviously. Since $p_1 > 0$, \eqref{eq:L(1-L)+A>=0} follows from
\[
 	L(x) (B(x)-L(x+1)) + A(x) 
 	= 
    \frac{p_1}x + O(x^{-2}).
\]
Finally, we show the inequality~\eqref{eq:L/S<1-1/x<(1-S-L)/S}. Since
\[
	\frac{L(x)}{S(x+1)} 
    = \frac{2/3 - s_1/(x+1)}{1/3 + s_1/(x+1)} \left( \frac12 + \frac{l_1}x \right) 
	=
	1 + \left( 2 l_1 - \frac92 s_1 \right) \frac1x + O(x^{-2}),
\]
the left-hand inequality in \eqref{eq:L/S<1-1/x<(1-S-L)/S} holds for all sufficiently negative $l_1$.
Since
\[
	\frac{B(x) - L(x+1)}{S(x+2)} = \frac{1-S(x+2)}{S(x+2)} \left( \frac12 - \frac{l_1}{x+1} \right) 
	=
	1 + \left( - 2 l_1 - \frac92 s_1 \right) \frac1x + O(x^{-2}),
\]
again, the right-hand inequality in \eqref{eq:L/S<1-1/x<(1-S-L)/S} holds for all sufficiently negative $l_1$.
\end{proof}

\subsection{Proof of Theorems~\ref{thm:NonPositive} and \ref{thm:Optimity}}

To show \cref{thm:NonPositive,thm:Optimity},
we need to construct recurrences that do not admit a (robustly) ultimately positive solution. 
Below, we construct such recurrences as a recurrence satisfied by three linearly independent sequences of first order. 
These three sequences form a basis of the solution space of the recurrence.
We use first-order sequences whose linear combinations are never (robustly) ultimately positive. 

\begin{lemma} \label{lem:No_positive_solution_generated}
	$i := \sqrt{-1}$. For $\alpha \in \R \setminus \{ 0 \}$, $\beta \in (-1, \frac12 )$ and $\gamma, \delta \in \R$, define
	\begin{equation} \label{eq:3sequences}
	g(n) := \prod_{k=1}^n  \left( \frac12 + \frac{1 + i \alpha}{4 k} \right),
	\quad\quad
	\overline{g}(n) := \overline{g(n)} ,
	\quad\quad
	h(n) := \prod_{k=1}^n \left( \beta + \frac{\gamma}k + \frac{\gamma + \delta}{k^2} \right).
\end{equation}
	Then any $\R$-linear combination of the real-valued sequences $g + \overline{g}$, $i(g - \overline{g})$ and $h$ is not robustly ultimately positive, where the robustness is with respect to the perturbations of coefficients of the linear combination.
    If $\beta + \frac{\gamma}x + \frac{\delta}{x^2} \leq 0$ holds additionally, then any $\R$-linear combination of those sequences is not ultimately positive.
\end{lemma}

The coefficient of $k^{-2}$ in the factor of $h$ is $\gamma + \delta$ rather than $\delta$ to simplify the calculations in the proof of the next lemma.

\begin{proof}
	Since $\arg \left( \frac12 + \frac{1 + i \alpha}{4 k} \right) = \arctan \left(  \frac{\alpha}{2 k} + O(k^{-2})  \right) = \frac{\alpha}{2 k} + O(k^{-2})$, as $n \to \infty$, the argument $\arg g(n)$ goes around the interval $[0, 2 \pi)$ infinitely many times while it slows down.
	
	Take any $\R$-linear combination $f(n) = s(g(n) + \overline{g}(n)) + t i (g(n) - \overline{g}(n)) + u h(n)$. We have $f(n) = 2s \re g(n) - 2t \im g(n) + u h(n)$.
    First, assume 
	$\beta + \frac{\gamma}x + \frac{\delta}{x^2} > 0$.
    Since
    $0 \leq \beta < \frac12$ now, $\lim\limits_{n \to \infty} \lvert h(n)\rvert / \lvert g(n)\rvert = 0$.
    By the observation on the argument, 
    $2s \re g(n), - 2t \im g(n) \leq 0$ and $\lvert \re g(n) \rvert / \lvert g(n) \rvert , \lvert \im g(n) \rvert / \lvert g(n) \rvert > \frac13$ for infinitely many $n$.
    Hence $f(n) < 0$ for infinitely many $n$ if $(s, t) \neq (0, 0)$. 
    Thus, if $f$ is ultimately positive, we have $s=t=0$. In this case, the ultimate positivity is not robust.
    
	Next, assume 
	$\beta + \frac{\gamma}x + \frac{\delta}{x^2} \leq 0$.
    Then
    $\beta + \frac{\gamma}x + \frac{\gamma + \delta}{x^2} \leq 0$ also holds, and we have $u h(2n) \leq 0$ or $u h(2n+1) \leq 0$ for sufficiently large $n$. 
    Without loss of generality, 
    we can assume $uh(2n) \leq 0$.
    By the observation on the argument, $2s \re g(2n), - 2t \im g(2n) \leq 0$ for infinitely many $n$.
    Thus $f(n) \leq 0$ for infinitely many $n$.
\end{proof}

In fact, under the condition of $\alpha, \gamma, \delta \in \R$, $\beta > -1$ and $\beta + \frac{\gamma}{k} + \frac{\gamma+\delta}{k^2} \neq 0$ for all $k \in \N_{>0}$, 
any $\R$-linear combination of $g + \overline{g}$, $i(g - \overline{g})$ and $h$ is not robustly ultimately positive if and only if $\alpha \neq 0 \land \beta + \frac{\gamma}x < \frac12 + \frac1{4x}$ or $\alpha=0 \land \beta + \frac{\gamma}x < -\frac12 - \frac1{4x}$.
We do not use this equivalence, so we proved a weaker one. 

Let us prove \cref{thm:Optimity}.
At the same time, we prove the next lemma, which will be used to prove \cref{thm:NonPositive}.

\begin{lemma}\label{lem:NonPositive}
For any $p_0, p_1, p_2 \in \R$ satisfying 
$p_0 + \frac{p_1}x + \frac{p_2}{x^2} \geq -\frac14 - \frac1{16 x^2}$
and any $\varepsilon > 0$, there exist $P \in \R(x)$ with $P(x) = p_0 + \frac{p_1}x + \frac{p_2}{x^2} + O(x^{-3})$ and $Q \in \R(x)$ with $Q(x) \geq R_P(x) - \frac{\varepsilon}{x^2}$ and $0 \notin Q(\N)$
such that the recurrence~\eqref{eq:3rd-order} does not admit an ultimately positive solution. 
Here, $R_P$ is the one in \eqref{item:-1/3<p0<=1} of  \cref{thm:positive_solution}.
\end{lemma}

\begin{proof}[Proof of \cref{thm:Optimity,lem:NonPositive}]
    Take $\alpha, \beta, \gamma, \delta \in \R$ and the sequences $g$, $\overline{g}$, $h$ as in    \cref{lem:No_positive_solution_generated}.
    We first construct a common third-order recurrence among these sequences.
    Let $\mathbf{N}$ be a shift operator such that $(\mathbf{N} f) (n) := f(n+1)$.
    Let $G(x)$ denote $\frac12 + \frac{1 + i \alpha}{4x} $.
    Then $g$ and $\overline{g}$ are annihilated by
	\[
		\left( \mathbf{N} - \frac{\overline{G(n+2)} - G(n+2)}{\overline{G(n+1)} - G(n+1)} \ \overline{G(n+1)} \right) \left( \mathbf{N} - G(n+1)  \right)
		=:
		\mathbf{N}^2 - \mathbf{N} - \widetilde{G}(n),
	\]
	where
    \[
		\widetilde{G}(x) =  -\frac14 - \frac{1 + \alpha^2}{16 (x+1)(x+2)} = -\frac14 - \frac{1 + \alpha^2}{16 x^2} + O(x^{-3}).
	\]
    Let $H(x)$ denote $\beta + \frac{\gamma}x + \frac{\gamma + \delta}{x^2}$. Then $g$, $\overline{g}$ and $h$ are all annihilated by 
    \begin{align}
	\label{eq:Anihilator}
		&\left( \mathbf{N} - \frac{H(n+3) H(n+2) - H(n+2) - \widetilde{G}(n+1)}{H(n+2) H(n+1) - H(n+1) - \widetilde{G}(n)} H(n+1) \right) \left( \mathbf{N}^2 - \mathbf{N} - \widetilde{G}(n) \right)
		\\
	\notag
		& \quad =:
		\mathbf{N}^3 - P_2(n) \mathbf{N}^2 - P_1(n) \mathbf{N} - P_0(n).
	\end{align}

    Next, we want to normalize this recurrence into the form of \cref{eq:3rd-order}.
    By $\lim\limits_{x \to \infty} P_2(x) = 1 + \beta > 0$, we have $P_2 > 0$.
	By normalization in \cref{sec:transformation},
    the sequences $g(n) \prod_{k=-2}^{n-3}P_2(k)^{-1}$, $\overline{g(n)} \prod_{k=-2}^{n-3}P_2(k)^{-1}$ and $h(n) \prod_{k=-2}^{n-3}P_2(k)^{-1}$
    satisfy the recurrence~\eqref{eq:3rd-order}.
    The coefficient $P$ has the following Taylor expansion at $x=\infty$.
    \[
		P(x) = \widetilde{p_0}(\beta) + \frac{\widetilde{p_1}(\beta, \gamma)}x + \frac{\widetilde{p_2}(\alpha, \beta, \gamma, \delta)}{x^2} + O(x^{-3}) 
	\]
    In this expansion, we have $\widetilde{p_0}(\beta) = -{(1 + \beta)^{-2}} \left( \frac14 +  \beta  \right)$, $\widetilde{p_1} \in \R(\beta)[\gamma]$ and $\widetilde{p_2} \in  \R(\beta)[\alpha, \gamma, \delta]$.
    Moreover, $\widetilde{p_1}$ and $\widetilde{p_2}$ are linear in $\gamma$ and $\delta$, respectively.

    Let us first assume $\alpha=0$ and show that 
    $(\widetilde{p_0}(\beta), \widetilde{p_1}(\beta, \gamma), \widetilde{p_2}(0, \beta, \gamma, \delta))$ ranges over $(-\frac13, \infty) \times \R \times \R$
    as 
    $(\beta, \gamma, \delta)$ varies over $(-1, \frac12 ) \times \R \times \R$.
    $\widetilde{p_0}( \beta ) = -{(1 + \beta)^{-2}} \left( \frac14 +  \beta  \right)$
    ranges over $( -\frac13, \infty )$ as $\beta$ varies.
    The coefficient of $\gamma$ in    
    $\widetilde{p_1}(\beta, \gamma)$ is $-(1+\beta)^{-3} (\frac12-\beta) \neq 0$, so $\widetilde{p_1}$ ranges over $\R$ as     
    $\gamma$ varies.
    Likewise, the coefficient of $\delta$ in $\widetilde{p_2}(\alpha, \beta, \gamma, \delta)$ is $-(1+\beta)^{-3} (\frac12 - \beta) \neq 0$, so $\widetilde{p_2}$ ranges over $\R$ as $\delta$ varies.

    Assume $\alpha = 0$ again, and calculate the Taylor expansions to check $Q(x) = R_P(x) + O(x^{-3})$.
    The result of the calculation is displayed in \cref{sec:TaylorCalculation}.
    (In this calculation, the coefficient of $x^{-2}$ is especially complicated. Since it is a quadratic polynomial in $\gamma$, 
    one possible way to calculate is to check each of the coefficients of $\gamma^2$, $\gamma^1$ and $\gamma^0$. You may use a computer to check the calculation.)

    Now, we are ready to show \cref{thm:Optimity}.
    For each $p_0$, $p_1$, $p_2$ and $\varepsilon$, let us construct suitable $P$ and $Q$. 
    By the above discussion, there exist $\beta \in \left( -1,\frac12 \right) $ and $\gamma, \delta \in \R$ such that 
    $(\widetilde{p_0}(\beta), \widetilde{p_1}(\beta, \gamma), \widetilde{p_2}(0, \beta, \gamma, \delta)) = (p_0, p_1, p_2)$.
    In the Taylor expansion of $P(x)$ and $Q(x)$ constructed above, the coefficients of $x^0$ and $x^{-1}$ do not depend on $\alpha$ or $\delta$. The coefficient of $x^{-2}$ is a polynomial (and in particular is continuous) in $\alpha$ and $\delta$.
    Moreover, $\widetilde{p_2}$ is linear in $\delta$ but not constant.
    Therefore, by a simultaneous perturbation of $\alpha$ and $\delta$,
    we can make $\alpha$ nonzero while keeping the value of $\widetilde{p_2}$ and the condition $Q(x) \geq R_P(x) - \frac{\varepsilon}{x^2}$.
    In addition, we can choose a perturbation such that rational functions used in the above construction do not have poles or zeros in $\N$ or in other suitable subsets of $\Z$.
    Then, these $P$ and $Q$ satisfy the condition required in the theorem by the former statement of    \cref{lem:No_positive_solution_generated}.

    We use a similar discussion for proving \cref{lem:NonPositive}.
    First, 
    under the assumption $\alpha = 0$, we show that $(\widetilde{p_0}(\beta), \widetilde{p_1}(\beta, \gamma), \widetilde{p_2}(0, \beta, \gamma, \delta))$ varies over $\{ (p_0, p_1, p_2) \mid p_0 + \frac{p_1}x + \frac{p_2}{x^2} \geq -\frac14 - \frac1{16 x^2} \}$ as $\beta \in (-1, \frac12 )$ and $\gamma, \delta \in \R$ vary under the condition $\beta + \frac{\gamma}x + \frac{\delta}{x^2} \leq 0$.
    $\widetilde{p_0}( \beta ) = -{(1 + \beta)^{-2}} \left( \frac14 +  \beta  \right)$ varies over the interval $[ -\frac14, \infty )$. If $\beta < 0$, then $\gamma$ and $\delta$ vary over $\R$, and a discussion similar to the above one works. If $\beta = 0$, then $\gamma$ varies over $( -\infty, 0]$, so $\widetilde{p_1}(0, \gamma) = - \frac{\gamma}2$ varies over $[0, \infty)$. Again, if $\gamma < 0$, we can use a similar discussion. If $\gamma = 0$, then $\delta$ varies over $( -\infty, 0]$, so $\widetilde{p_2}(0, 0, 0, \delta) = -\frac{\delta}2 - \frac1{16}$ varies over $[-\frac1{16}, \infty)$.
    Next, by a simultaneous perturbation of $\alpha$ and $\delta$,
    we want to make $\alpha$ nonzero while keeping the value of $\widetilde{p_2}$ and the condition $Q(x) \geq R_P(x) - \frac{\varepsilon}{x^2}$.
    If this is possible, the lemma follows from the latter statement of \cref{lem:No_positive_solution_generated}.
    The above discussion will do unless $\beta=\gamma=\delta=0$. 
    In the case $\beta=\gamma=\delta=0$, possible perturbations of $\delta$ are only in the negative direction;
    since $\widetilde{p_2}(\alpha, 0, 0, \delta) = -\delta/2 - (1+\alpha^2)/16$, such a perturbation will do.
\end{proof}

Even if we use a stronger version (equivalence version) of \cref{lem:No_positive_solution_generated}, we cannot obtain any $P$ such that $P(x) = -\frac13 + \frac{p_1}x + O(x^{-2}) \ (p_1 > 0)$ through the above construction. 
Because of this problem, we failed to show the necessity of \cref{thm:positive_solution} in case~\eqref{item:p0=-1/3_p1>0}.
For case \eqref{item:p0=-1/3_p1=0}, we can obtain some results like \cref{thm:Optimity}, but these results do not cover the whole case.

\begin{proof}[Proof of \cref{thm:NonPositive}]
The existence of an ultimately positive solution of the recurrence~\eqref{eq:3rd-order} follows from \cref{thm:positive_solution} if
$Q(x) \succeq_3 R_P(x)$.
Let us show the inverse.
Assume $Q(x) \prec_3 R_P(x)$.
$R_P(x) \preceq_3 0$ follows from $P(x) \prec_3 0$ and the inequality~\eqref{eq:<=>exists_2nd-order_posi_sol}. Thus $Q \prec_3 0$, and so $Q<0$. 

Under the condition of $P, Q \leq 0$, the recurrence~\eqref{eq:3rd-order} has an ultimately positive solution if and only if there exists an ultimately positive sequence $S$ such that
\begin{equation}\label{eq:S>=1+P/S+Q/SS}
	S(n+2) \leq 1 + \frac{P(n)}{S(n+1)} + \frac{Q(n)}{S(n+1)S(n)}
\end{equation}
for all sufficiently large $n$, which you can verify as follows.
If it admits an ultimately positive solution $f$, then $S(n) := f(n+1) / f(n)$ is a desired sequence.
To show the converse, assume the existence of $S$. 
The right-hand side of the inequality~\eqref{eq:S>=1+P/S+Q/SS} is monotonically increasing in $S(n)$, $S(n+1)$, so there is also an ultimately positive sequence $S$ such that, for $n$ large enough,
\[
	S(n+2) = 1 + \frac{P(n)}{S(n+1)} + \frac{Q(n)}{S(n+1)S(n)}.
\]
Then, for some large $N \in \N$, we can take a solution $f$ such that $f(n) = \prod_{k=N}^{n-1} S(k)$ for all $n \geq N$. This is an ultimately positive solution.

Since $P, Q \prec_3 0$, by \cref{lem:NonPositive}, there exist $\widetilde{P}, \widetilde{Q} \in \R(x)$
such that 
$P \leq \widetilde{P} \leq 0$, $Q \leq \widetilde{Q}$, and the recurrence~\eqref{eq:3rd-order} with its coefficients replaced by $\widetilde{P}$ and $\widetilde{Q}$
does not admit an ultimately positive solution. 
(When we use \cref{lem:NonPositive}, set $p_0$, $p_1$ as the coefficients of $x^0$, $x^{-1}$ in the Taylor expansion of $P(x)$ and use $p_2$ that is larger than but sufficiently close to the coefficient of $x^{-2}$. Also, use a sufficiently small positive $\varepsilon$.)
By \cref{thm:positive_solution}, we have $\widetilde{Q}(x) \prec_3 R_{\widetilde{P}}(x)  \leq 0$.
Then, it follows from the above equivalence that there is no ultimately positive sequence $S$ that satisfies the inequality~\eqref{eq:S>=1+P/S+Q/SS} whose $P$ and $Q$ are replaced by $\widetilde{P}$ and $\widetilde{Q}$ respectively. 
Since this inequality is monotonically increasing in $P$ and $Q$, 
there is no ultimately positive sequence $S$ that satisfies the original version of the inequality~\eqref{eq:S>=1+P/S+Q/SS}.
Again by the above equivalence, the recurrence~\eqref{eq:3rd-order} does not admit an ultimately positive solution.
\end{proof}

\section{Application}
\label{sec:execution}

In this section, we apply our \cref{proc:overview} using the positivity certificate~\eqref{eq:positivity_certificate} to specific third-order sequences.

The coefficient condition of \cref{thm:procedure} is just for obtaining the characterization of termination. Actually, our procedure works if we can construct appropriate $S, L \in (\overline{\Q} \cap \R) (x)$.
Such $S$, $L$ can be constructed whenever $P$ and $Q$ satisfy one of the conditions in \cref{thm:positive_solution}.
Let us apply our procedure to $P$, $Q$, $f$ in \cref{ex:2f1<f2}.

\begin{example} \label{ex:auto-proof}
Let $P$, $Q$, $f$ be as in \cref{ex:2f1<f2}, $S$ as in \eqref{eq:construction_of_S} with $s_2 = 0$, and $L := \left( \frac12 + \frac1{4x} \right) B(x-1)$.
Let us first compute $N \in \N$ in line~\ref{state:findN} of \cref{proc:overview}.
According to the proof of \cref{lem:sufficient_condition},
it suffices to choose $N$ such that
the conditions~\eqref{eq:0<=S<1} and \eqref{eq:C>=0} of \cref{lem:QE} and \eqref{eq:0<L<=B} and ~\eqref{eq:L(1-L)+A>=0} of \cref{lem:2nd-order}
literally hold for all $x \geq N-1$.
Thus, we can check that $N=2$ will do by using a computer. (Note that $S(0)$ is undefined.)
Then decide whether $\cert(2)$, $\cert(3)$, \dots hold and whether $f(0) > 0$, $f(1)>0$, \dots. We find that $\cert(6)$ is true and $f(0), f(1), \dots, f(5) > 0$, which proves the positivity of $f$.
Note that, in the certificate, not the original $f$ but the normalized sequence $f(n) \prod_{k=-2}^{n-3} P_2(k)^{-1}$ is used.

The positivity of $f$ leads to the following inequality, with an additional verification for $n = 0, 1, 2, 3$.
\[
	2 n \sum_{k=0}^n \binom{n}{k}^3 <  8^n \quad (n=0, 1, 2, \dots)
\]
\end{example}

\begin{example} \label{ex:auto-proof2}
Let $\alpha$ be the unique negative number satisfying $\alpha^3 - 24 \alpha^2 + 3 \alpha + 1 = 0$. 
Then our procedure succeeds in proving the positivity of $f$ in \cref{ex:}.
Let $S$ be as in  \cref{eq:construction_of_S} with $s_2 =0$.
Because $L$ used in the proof is slightly complicated, we use $L(x) = \left( \frac12 + \frac1{4x} \right) B(x-1)$ instead.
This change does not affect the correctness of the procedure  as long as the conditions~\eqref{eq:0<=S<1},~\eqref{eq:C>=0} of \cref{lem:QE} and the two conditions in \cref{lem:2nd-order} hold.
With a computer, one can check that $N=2$ will do for $N$ in line~\ref{state:findN} of \cref{proc:overview} and that 
$\cert(2)$ holds.
Since $f(0) = 1>0$ and $f(1) = 2>0$, we conclude that $f$ is positive.
Note, as in \cref{ex:auto-proof}, that the normalized sequence $f(n) \prod_{k=-2}^{n-3} P_2(k)^{-1}$ is used in the certificate instead of $f$.
\end{example}

\section{Future work}

One limitation of this research is that the assumptions of \cref{thm:NonPositive} and \cref{thm:procedure}~\eqref{item:-1/3<=p0<=-1/4} are disjoint.
This is due to our technique, so we want to improve it using some new ideas. 
In particular, we want to show \cref{conj:}, and furthermore, want to obtain a necessary and sufficient condition for the existence of robustly ultimately positive solutions.

Our approach also
proves a higher-order version of \cref{thm:positive_solution}.

\begin{proposition}
	Let $r \geq 2$. 
    Assume that the recurrence~\eqref{eq:holonomic} is of Poincar\'{e} type and $0 \notin P_0(\N)$. Let $P_i(x) = p^{(i)} + q^{(i)} / x + O(x^{-2})$ be the Taylor expansion of $P_i$ at $x=\infty$ for each $i$.
    Assume that all of the roots $\lambda_1, \dots, \lambda_r$ of the characteristic polynomial~\eqref{eq:characteristic_polynomial}
    are real, and arrange them in order so that $\lambda_1 \geq \lambda_2 \geq \dots \geq \lambda_r$.
    Moreover, assume $\lambda_{r-1} \geq 0$. 
    Let 
	$\psi_i(\lambda)$ denote $q^{(r-1)} \lambda^{i} +  q^{(r-2)} \lambda^{i-1} + \dots + q^{(r-1-i)} \lambda^{0}$.
    Then, if one of the following is satisfied, the recurrence~\eqref{eq:holonomic} admits a robustly ultimately positive solution.
	\begin{enumerate}[(i)]
	\item \label{item:lambda1}
	$\lambda_r \geq -\lambda_2$ and $\psi_i(\lambda_1) > 0$ for all $i=0, 1, \dots, r-1$.
	\item \label{item:lambda2}
	$- \lambda_2 > \lambda_r \geq -\lambda_1$ and $\psi_i(\lambda_2) > 0$ for all $i=0, 1, \dots, r-1$.
	\end{enumerate}
\end{proposition}

\begin{proof}
Let us first assume $r \geq 3$.
Set $\lambda = \lambda_1$ in the case~\eqref{item:lambda1}, and $\lambda = \lambda_2$ in the case~\eqref{item:lambda2}.
Choose $\lambda'$ such that $\{ \lambda_1, \lambda_2 \} = \{ \lambda, \lambda' \}$.
Let $\mu := \frac12 ( \lambda_r + \lambda') \ ( \geq 0 )$.
For $r \geq 4$, we use the positivity certificate
\[
	\cert(n) = 
	\quad
	\begin{aligned}
	f(n) &> 0 \ \land \ \Delta_{\lambda} f(n) > 0 \ \land \ \Delta_{\lambda_3}\Delta_{\lambda} f(n) > 0
	\ \land \cdots
	\\
	& \land \ 
	\Delta_{\lambda_{r-1}} \dots \Delta_{\lambda_3}\Delta_{\lambda} f(n) > 0
	\ \land \ \Delta_{\mu} \Delta_{\lambda_{r-1}} \dots \Delta_{\lambda_3}\Delta_{\lambda} f(n) > 0.
	\end{aligned}
\]
For $r = 3$, we use $\cert(n) = f(n) > 0 \ \land \ \Delta_{\lambda} f(n) > 0 \land  \ \Delta_{\mu} \Delta_{\lambda} f(n) > 0$.
It suffices to show that, for any solution $f$ of the recurrence~\eqref{eq:holonomic}, $\cert(n) \implies \cert(n+1)$
for all sufficiently large $n$, 
since we obtain a robustly ultimately positive solution $f$ by choosing such a large $n$ and an initial value such that $\cert(n)$ holds.

Let $S_i(x_1, \dots, x_j)$ be the elementary symmetric polynomial of degree $i$ in 
$x_1$, \dots, $x_j$. Since $p^{(i)} = (-1)^{r-1-i} S_{r-i}(\lambda_1, \dots, \lambda_r)$, we can rewrite the recurrence~\eqref{eq:holonomic} in the following form.
\begin{align*}
	\Delta_{\lambda} f(n+r-1) &= (P_{r-1}(n) - \lambda) \Delta_{\lambda} f(n+r-2) + \cdots 
	\\
	& \quad +
	(P_{1}(n) + \dots + P_{r-1}(n) \lambda^{r-2} - \lambda^{r-1}) \Delta_{\lambda} f(n)
	\\
	& \quad + (P_0(n) + P_1(n) \lambda + \dots + P_{r-1} (n) \lambda^{r-1} - \lambda^r) f(n)
	\\
	&= (-1)^0 S_1(\lambda', \lambda_3, \dots, \lambda_r) \Delta_{\lambda} f(n+r-2) + \cdots 
    \\
	& \quad +
	(-1)^{r-2} S_{r-1}(\lambda', \lambda_3, \dots, \lambda_r) \Delta_{\lambda} f(n)
	\\
	& \quad 
	+ 
	\left( \frac{\psi_{0}(\lambda)}n + O(n^{-2}) \right)\Delta_{\lambda} f(n+r-2)
	+ \dots 
    \\
    & \quad 
    +	\left( \frac{\psi_{r-2}(\lambda)}n + O(n^{-2}) \right)
	\Delta_{\lambda} f(n) 
	+
	\left( \frac{\psi_{r-1}(\lambda)}n + O(n^{-2}) \right) f(n).
\end{align*}
For the second equation, we used $p^{(0)} \lambda^0 + \dots + p^{(r-1)} \lambda^{r-1} - \lambda^r = 0$, which is derived from the fact that $\lambda$ is a root of the characteristic polynomial. 
By the assumption, all $\frac{\psi_{i}(\lambda)}n + O(n^{-2}) \ (i=0, 1, \dots, r-1)$ are positive for all sufficiently large $n$.
For such $n$, we prove $\cert(n) \implies \cert(n+1)$. Let us assume $\cert(n)$.
The components other than $\Delta_{\mu} \Delta_{\lambda_{r-1}} \dots \Delta_{\lambda_3}\Delta_{\lambda} f(n+1) >0 $ in $\cert(n+1)$ immediately follow from $\cert(n)$, so we show this inequality. By the above equation and $\cert(n)$, we have
\begin{align*}
    \Delta_{\lambda} f(n+r-1) 
    > 
    (-1)^0  S_1(\lambda', \lambda_3, \dots, \lambda_r) & \Delta_{\lambda} f(n+r-2) + \cdots 
    \\
	&+
	(-1)^{r-2} S_{r-1}(\lambda', \lambda_3, \dots, \lambda_r) \Delta_{\lambda} f(n).
\end{align*}
Since the coefficients of $\Delta_{\lambda} f(n+i)$ are again elementary symmetric polynomials, we can inductively transform the inequality as below.
\[
	\Delta_{\lambda_{r-1}} \dots \Delta_{\lambda_3}\Delta_{\lambda} f(n+2) 
	>
	(\lambda_r + \lambda') \Delta_{\lambda_{r-1}} \dots \Delta_{\lambda_3}\Delta_{\lambda} f(n+1)
	- \lambda_r \lambda' \Delta_{\lambda_{r-1}} \dots \Delta_{\lambda_3}\Delta_{\lambda} f(n).
\]
This gives 
\begin{align*}
    \Delta_{\mu} \Delta_{\lambda_{r-1}} \dots \Delta_{\lambda_3}\Delta_{\lambda} f(n+1) 
	>
	\mu \Delta_{\mu} \Delta_{\lambda_{r-1}} &\dots \Delta_{\lambda_3}\Delta_{\lambda} f(n)
    \\
	&+
    ( \mu^2 - \lambda_r \lambda') \Delta_{\lambda_{r-1}} \dots \Delta_{\lambda_3}\Delta_{\lambda} f(n).
\end{align*}
Since
$\mu^2 - \lambda_r \lambda' = \frac14 (\lambda_r - \lambda')^2 \geq 0$,
we have $\Delta_{\mu} \Delta_{\lambda_{r-1}} \dots \Delta_{\lambda_3}\Delta_{\lambda} f(n+1) >0 $.

In the case $r=2$, 
we use $\cert(n) = f(n) > 0 \ \land \ \Delta_{\mu} f(n) > 0$ and $\mu = \frac12 (\lambda_1 + \lambda_2)$.
Since we have 
\[
    \Delta_{\mu} f(n+1) = \left(\mu+\frac{q^{(1)}}n + O(n^{-2})\right) \Delta_{\mu} f(n)+\left(\mu^2 - \lambda_1 \lambda_2 + \frac{q^{(1)}\mu + q^{(0)}}n + O(n^{-2})\right) f(n),
\]
a similar discussion leads to the desired conclusion.
\end{proof}

This proposition handles some cases with several or multiple dominant roots for general-order recurrences, and thus partially generalizes \cref{thm:simple_unique_posi_sol}.
However, we have not found a necessity result like \cref{thm:NonPositive,thm:Optimity}.
This proposition could be generalized much further.
It is also worth exploring ways to generalize \cref{thm:procedure} to the case of order four or more.

\bibliography{Bib_Positivity_Proving20261107}

\appendix

\section{Omitted Taylor expansions in Example~\ref{ex:}}
\label{sec:SpecificCoefficients}

$P_2$, $P_1$ and $P_0$ in the case $t=\alpha$ in \cref{ex:} are the following rational functions:
\begin{align*}
	P_2(x) &= \frac{(\alpha^2 - 25 \alpha + 25)(333 x^2  - (24 \alpha^2   - 384 \alpha - 1557 )x - 52 \alpha^2  + 832 \alpha + 1838)}{666 (x+3)(2x+5)},
	\\
	P_1(x) &= \frac{(2\alpha^2 - 59 \alpha + 17)(-543 x  + 26 \alpha^2   - 632 \alpha - 911)(x+2)}{543 (x+3)(2x+5)}, 
	\\
	P_0(x) &= \frac{ \alpha (4 \alpha - 1)(x+1)(x+2)}{2 (x+3)(2x+5)}.
\end{align*}
The Taylor expansions are:
\begin{align*}
	P_2(x) &= \frac{\alpha^2 - 25 \alpha + 25}{4} - \frac{11 \alpha^2 - 275 \alpha + 131}{24 x} + \frac{71 \alpha^2 - 1775 \alpha + 815}{48 x^2} + O(x^{-3}),
	\\
	P_1(x) &= -\frac{2\alpha^2 - 59 \alpha + 17}{2} + \frac{26 \alpha^2 - 713 \alpha + 203}{12 x} - \frac{170 \alpha^2 - 4637 \alpha + 1319}{24 x^2} + O(x^{-3}), 
	\\
	P_0(x) &= \alpha^2 - \frac14 \alpha - \frac{5 \alpha ( 4 \alpha - 1) }{8 x} + \frac{33 \alpha( 4 \alpha - 1)}{16 x^2} + O(x^{-3}).
	\end{align*}
The Taylor expansions of $P$ and $Q$ are
	\begin{align*}
		P(x) =& \frac{1296}{1369} \alpha^2 + \frac{576}{1369} \alpha - \frac{856}{4107} 
		+\left(  \frac{12111832}{455877} \alpha^2 - \frac{1533760}{455877} \alpha -  \frac{505988}{455877} \right)\frac1x 
		\\
		& \quad +\left(  \frac{24740295424}{50602347} \alpha^2 - \frac{3149159440}{50602347} \alpha -  \frac{1038656716}{50602347} \right)\frac1{x^2} 
		+ O(x^{-3}),
		\\
		Q(x) =& -\frac{459648}{50653} \alpha^2 + \frac{58560}{50653} \alpha + \frac{520048}{1367631}
        \\
		&+\left(  -\frac{18646907392}{50602347} \alpha^2 + \frac{2376005056}{50602347} \alpha +  \frac{781102736}{50602347} \right)\frac1x 
		\\
		& \quad +\left(  -\frac{5886260020192}{624095613} \alpha^2 + \frac{750038675920}{624095613} \alpha +  \frac{2219130104972}{5616860517} \right)\frac1{x^2} 
		+ O(x^{-3}).
	\end{align*}
The rational function $R_P$ is
	\begin{align*}
	R_P(x) = & -\frac{459648}{50653} \alpha^2 + \frac{58560}{50653} \alpha + \frac{520048}{1367631}
    \\
		&+\left(  -\frac{18646907392}{50602347} \alpha^2 + \frac{2376005056}{50602347} \alpha +  \frac{781102736}{50602347} \right)\frac1x 
		\\
		& \quad +\left(  -\frac{52978492293536}{5616860517} \alpha^2 + \frac{6750667684544}{5616860517} \alpha +  \frac{739735883536}{1872286839} \right)\frac1{x^2}.
	\end{align*}
	Using a computer, we can check $Q(x) = R_P(x) + O(x^{-2})$ and $Q \succ_3 R_P$.

\section{Calculations omitted in the proof of Theorem~\ref{thm:Optimity} and Lemma~\ref{lem:NonPositive}}
\label{sec:TaylorCalculation}

In this section, we display the Taylor expansions of $P$ and $Q$ at $x=\infty$, which were omitted in the proof of 
\cref{thm:Optimity,lem:NonPositive}.
The denominator of the fraction in the left factor of the operator~\eqref{eq:Anihilator} is expanded as 
    \[
	H(x+2) H(x+1) - H(x+1) - \widetilde{G}(x) = \left( \frac12 - \beta \right) \left(  \frac12 - \beta - \frac{2 \gamma}{x} \right) + O(x^{-2}).
	\]
Then, since $\beta < \frac12$, the fraction is expanded as $1 + \frac{4\gamma}{(1-2 \beta)x^2}  + O(x^{-3})$.
Multiplying this by
$H(x+1) = \beta + \frac{\gamma}x + \frac{\delta}{x^2} + O(x^{-3})$,
we obtain $\beta + \frac{\gamma}x + \frac{\delta}{x^2} + \frac{4 \gamma \beta}{(1-2 \beta) x^2} + O(x^{-3})$.
Thus 
\begin{align*}
		P_2(x) 
		&= 
		1 + \beta + \frac{\gamma}x + \frac{\delta}{x^2} + \frac{4\gamma \beta}{(1-2 \beta) x^2} + O(x^{-3}),
		\\
		P_1(x) &= - \frac14 - \beta - \frac{\gamma}x - \frac{\delta}{x^2} - \frac{4\gamma \beta}{(1-2 \beta) x^2}  - \frac{1 + \alpha^2}{16 x^2}  + O(x^{-3}),
		\\
		P_0(x) &= \frac14 \left(  \beta + \frac{\gamma}x + \frac{\delta}{x^2} + \frac{4\gamma \beta}{(1-2 \beta) x^2}  \right) + \frac{\beta ( 1 + \alpha^2)}{16 x^2} + O(x^{-3}),
	\end{align*}
and so
    \[
	P_2(x)^{-1} = \frac{1}{(1 + \beta)^2} \left( 1 + \beta - \frac{\gamma}x - \frac{\delta}{x^2} + \frac{\gamma}{x^2} \left( \frac{\gamma}{1+\beta} - \frac{4 \beta}{1-2\beta} \right)   \right) + O(x^{-3}).
	\]
It follows that 
	\begin{align*}
		P(x) &= 
		\frac{1}{(1 + \beta)^4}
		\left( 
		(1 + \beta)^2 - \frac{2 (1+\beta) \gamma}x + \frac{3\gamma^2 - (\gamma + 2 \delta) (1 + \beta)}{x^2} - \frac{8 \beta \gamma(1+\beta) }{(1-2\beta) x^2} \right) 
		P_1(x) 
        \\
        & \quad\quad + O(x^{-3}),
		\\
		Q(x) &= 
		\frac{1}{(1 + \beta)^5} \left( (1+\beta)^2 - \frac{3 (1 + \beta) \gamma }x + \frac{6 \gamma^2 - 3 \delta (1+\beta)}{x^2} - \frac{3 \gamma (1+2 \beta) (1+\beta)}{(1-2\beta)x^2} \right) P_0(x) 
        \\
        & \quad\quad + O(x^{-3}).
	\end{align*}
Under the condition $\alpha=0$, a straightforward calculation gives the desired result:
\begin{align*}
&\begin{aligned}
	P(x) = &- \frac{1/4 + \beta}{(1+\beta)^2} - \frac{(1/2 - \beta) \gamma}{(1+\beta)^3 x}
	+ \frac{(5/4 - \beta) \gamma^2}{(1+\beta)^4 x^2}
	\\
    &+ \left(  \left( \frac14 -  \beta \right) \gamma - \frac{1+\beta}{16} - \frac12 (1 - 2 \beta) \delta \right) \frac1{(1+\beta)^3 x^2} + O(x^{-3})
\end{aligned}
\\
&\begin{aligned}
	Q(x) = & \frac{\beta}{4(1+\beta)^3} + \frac{(1/2 - \beta) \gamma}{2 (1+\beta)^4 x}
	- \frac{3(1 - \beta) \gamma^2}{4(1+\beta)^5 x^2}
	\\
    &+ \left(  \frac14 \beta \gamma + \frac{(1+\beta) \beta}{16} + \frac14 (1-2\beta) \delta \right) \frac1{(1+\beta)^4 x^2} + O(x^{-3}).
\end{aligned}
\end{align*}

\end{document}